\documentclass[conf]{new-aiaa}
\usepackage[utf8]{inputenc}

\usepackage{graphicx}
\usepackage{amsmath}
\usepackage[version=4]{mhchem}
\usepackage{siunitx}
\usepackage{longtable,tabularx}
\usepackage{subfigure}
\usepackage{float}
\usepackage{amsfonts}
\usepackage[T1]{fontenc}
\usepackage{xcolor}
\usepackage{soul}
\usepackage{algorithm}
\usepackage{algorithmic}
\usepackage{caption}
\usepackage[capitalize]{cleveref}
\def\foo ABC{DGRM}

\title{Decentralized Graph-Based Multi-Agent Reinforcement Learning Using Reward Machines}

\author[1]{Jueming Hu}

\author[1]{Zhe Xu}

\author[1]{Weichang Wang}

\author[2]{Guannan Qu}

\author[1]{Yutian Pang}

\author[1]{Yongming Liu}

\affil[1]{Arizona State University}
\affil[2]{Carnegie Mellon University}

\begin{document}

\maketitle
\begin{abstract}

In multi-agent reinforcement learning (MARL), it is challenging for a collection of agents to learn complex temporally extended tasks. The difficulties lie in computational complexity and how to learn the high-level ideas behind reward functions. We study the graph-based Markov Decision Process (MDP) where the dynamics of neighboring agents are coupled. To learn complex temporally extended tasks, we use a reward machine (RM) to encode each agent's task and expose reward function internal structures. RM has the capacity to describe high-level knowledge and encode non-Markovian reward functions. We propose a decentralized learning algorithm to tackle computational complexity, called decentralized graph-based reinforcement learning using reward machines (\foo ABC), that equips each agent with a localized policy, allowing agents to make decisions independently, based on the information available to the agents. \foo ABC uses the actor-critic structure, and we introduce the tabular Q-function for discrete state problems. We show that the dependency of Q-function on other agents decreases exponentially as the distance between them increases. Furthermore, the complexity of \foo ABC is related to the local information size of the largest $\kappa$-hop neighborhood, and \foo ABC can find an $O(\rho^{\kappa+1})$-approximation of a stationary point of the objective function. To further improve efficiency, we also propose the deep \foo ABC algorithm, using deep neural networks to approximate the Q-function and policy function to solve large-scale or continuous state problems. The effectiveness of the proposed \foo ABC algorithm is evaluated by two case studies, UAV package delivery and COVID-19 pandemic mitigation. Experimental results show that local information is sufficient for \foo ABC and agents can accomplish complex tasks with the help of RM. \foo ABC improves the global accumulated reward by 119\% compared to the baseline in the case of COVID-19 pandemic mitigation.

\end{abstract}

\section{Introduction}

In multi-agent reinforcement learning (MARL), a collection of agents interact within a common environment and learn to jointly maximize a long-term reward. We study MARL in a graph-based Markov Decision Process (MDP) setting, where for each agent, the transition model is represented by a dynamic Bayesian network \cite{dean1989model, guestrin2003efficient}. In graph-based MDPs, the transition function between an agent's states may depend on its current state and action and the neighboring agents \cite{forsell2006approximate, cheng2013variational}. In this work, the agents are allowed to have different reward functions from different tasks, and they can perceive their own rewards.

The key challenge in MARL is the combinatorial nature, which results in high computational complexity. The combinatorial nature refers to the exponentially increased size of the joint state and action space in the multi-agent system \cite{blondel2000survey, hernandez2019survey}. Since all agents are learning simultaneously, the behavior of each individual agent relies on what has been learned by the respective other agents. Thus, the learning problem is non-stationary from each agent's perspective. To tackle non-stationarity, each individual agent has to consider the joint state and action space, whose dimension increases exponentially with the number of agents.  

RL becomes more challenging when solving complex temporally extended tasks. In many complex tasks, agents only receive sparse rewards for complex behaviors over a long time. However, agents do not have access to the high-level ideas behind the sparse rewards. Moreover, in a sparse-reward formulation, if it is improbable that the agent achieves the task by chance when exploring the environment, the learning agent will rarely obtain rewards.


In this paper, we provide a decentralized framework that enables a collection of agents to solve complex temporally extended tasks under coupled dynamics with enhanced computational efficiency. Specifically, we use reward machines (RMs) to describe the environment, track the progress through the task, and encode a sparse or non-Markovian reward function for each agent. The proposed algorithm, called decentralized graph-based reinforcement learning using reward machines (\foo ABC),  uses the actor-critic structure, specifically adopting temporal difference (TD) learning to train the truncated Q-function and policy gradient for the localized policy. To tackle computational efficiency, we propose a truncated Q-function considering only the information of the $\kappa$-hop neighborhood. The RM state is augmented with the low-level MDP state as the input for the truncated Q-function and the policy function. Each agent is equipped with a localized policy, and information exchange among agents when implementing the learned policies is not required. We provide proofs that the truncated Q-function can approximate Q-function with high accuracy and the error of policy gradient approximation is bounded. Also, \foo ABC can find an $O(\rho^{\kappa+1})$-approximation of a stationary point of the objective function. To further improve efficiency, we develop deep \foo ABC which uses deep neural networks for function approximation to improve scalability. We demonstrate the effectiveness of \foo ABC and deep \foo ABC on a discrete state problem of UAV package delivery and a continuous state problem of COVID-19 pandemic mitigation, respectively. In the experiments, agents can accomplish complex tasks under the guidance of RMs. 

\section{Related Work}

Recent years have witnessed the rapid development in MARL \cite{littman1994markov, claus1998dynamics, littman2001value, hu2003nash}. An MARL system can be categorized into centralized and decentralized control. The central controller decides the actions for all agents \cite{silver1990ils,hu2020uas}, while within the decentralized control, agents make their own decisions based on their observations \cite{hu2020uas,wang20213m}. The central controller needs to collect information of all agents and communicate with all agents, which mitigates the non-stationarity but may decrease the scalability of the multi-agent system. Therefore, we consider a decentralized learning algorithm to solve typically graph-based MDP, where the dynamics of agents are coupled. Graph-based MDP can be also referred to as networked multi-agent MDP \cite{zhang2018fully}, factored MDP \cite{guestrin2001multiagent,guestrin2003efficient,cubuktepe2021distributed}. 

The most relevant studies to our proposed \foo ABC algorithm are the methods utilizing the actor-critic structure for MARL.
The authors in \cite{foerster2018counterfactual, lowe2017multi} develop decentralized actor-centralized critic models for MARL. The centralized critic has access to the joint action and all available state information. The decentralized actor considers the local information of each agent. Additionally, the proposed algorithms in \cite{foerster2018counterfactual, lowe2017multi} adopt the framework of centralized training with decentralized execution. After training is completed, only the local actors are used at the execution phase. The communication requirement in our proposed \foo ABC algorithm is weaker than that in \cite{foerster2018counterfactual, lowe2017multi}, where the localized policy only requires its own MDP state and RM state, and observations of other agents are not necessary. A decentralized actor-critic algorithm with provable convergence guarantees is proposed in \cite{zhang2018fully}. They use linear approximation for the Q-function, but it is unclear whether the loss caused by the function approximation is small. The proposed \foo ABC algorithm is inspired by the Scalable Actor Critic (SAC) algorithm in \cite{qu2020scalable} and we further propose the \foo ABC algorithm to solve complex temporally extended tasks and the deep \foo ABC algorithm for large-scale or continuous state problems.

Recently, reward machines (RMs) have received much attention for task specification. The authors in \cite{icarte2018using} propose reward machines, i.e., a type of Mealy machines, to encode structures or high-level knowledge behind the tasks. They develop the Q-learning for Reward Machines (QRM) algorithm in the single-agent setting and show that QRM can converge to an optimal policy in the tabular case. Later, they extend their work and develop counterfactual experiences for reward machines (CRM) and hierarchical RL for reward machines (HRM). 

In the multi-agent setting, the authors in \cite{neary2020reward} use RMs to describe cooperative tasks and introduce how to decompose the cooperative task into a collection of new RMs, each encoding a sub-task for an individual agent. They assume agents only observe their local state and abstracted representations of their teammates \cite{neary2020reward}. Agents in our work have access to the information of neighboring agents, since we consider the graph-based MDP, where the dynamics of neighboring agents are coupled. Thus, the dimension of Q-function in our work is larger than that in \cite{neary2020reward} and needs complexity reduction for better scalability. RMs have also been applied to robotics, such as quadruped locomotion learning \cite{defazio2021learning} and planning \cite{shah2020planning}. There are also studies focusing on learning RMs from experience instead of giving as a priori to the learning agent \cite{xu2020joint, toro2019learning,furelos2020induction,furelos2021induction,hasanbeig2021deepsynth,rens2020learning}. Hierarchical reinforcement learning (HRL) is another classical approach to solve complex tasks, which decomposes an RL problem into a hierarchy of subtasks, including methods such as HAMs \cite{parr1998reinforcement}, Options \cite{sutton1999between}, and MAXQ \cite{dietterich2000hierarchical}. However, these HRL approaches cannot guarantee convergence to the optimal policy.

\section{Preliminaries}

\subsection{Markov Decision Processes and Reward Machines}

A labeled MDP is defined by a tuple,  $\mathcal{M} = (S, s_I, A, R, P, \gamma, \mathcal{P}, L)$, where $S$ is a finite set of states;  $s_I \in S$ is an initial state; $A$ is a finite set of actions; $R:S \times A \to \mathbb{R}$ is the reward function, $R(s, a)$ represents the reward obtained at state $x$ after taking action $a$; $P: S \times A \times S \to [0, 1]$ is the transition function, $P(s, a, s')$ represents the probability of going to $s'$ from $s$ after taking action $a$; $\gamma$ is the discount factor, $\gamma \in (0, 1]$; $\mathcal{P}$ is a finite set of events; and $L: S \times A \times S \to 2^\mathcal{P}$ is a labeling function,  $L(s, a, s')$ maps the state transition to a relevant high-level event.

\textbf{Definition 1.} A reward machine (RM) \emph{is a tuple, $\mathcal{R} = (U, u_I, \mathcal{P}, \delta, \sigma)$, where $U$ is a finite set of RM states, $u_I \in U$ is an initial state, $\mathcal{P}$ is a finite set of events, $\delta$ is the transition function: $U \times 2^{\mathcal{P}} \to U$, and $\sigma$ is a reward function: $U \times 2^{\mathcal{P}} \to \mathbb{R}$.}

Reward machines \cite{icarte2020reward} are a way to encode a non-Markovian reward function. A run of a reward machine $\mathcal{R}$ on the sequence of labels $\ell_1 \ell_2 \dots \ell_k \in (2^{\mathcal{P}})^*$ is a sequence $u_0 (\ell_1, r_1) u_1 (\ell_2, r_2) \dots u_{k-1} (\ell_k, r_k) u_k$ of states and label-reward pairs such that $u_0 = u_I$. For all $i \in \{0, \dots, k-1 \}$, we have $\delta(u_i, l_{i+1}) = u_{i+1}$ and $\sigma(u_i, l_{i+1})=r_{i+1}$. We write $\mathcal{R}(\ell_1 \ell_2 \dots \ell_k) = r_1 r_2 \dots r_k$ to connect the input label sequence to the sequence of rewards produced by the machine $\mathcal{R}$. We say that a reward machine $\mathcal{R}$ encodes the reward function $R$ of an MDP if for every trajectory $s_0 a_1 s_1 \dots a_k s_k$ and the corresponding label sequence $\ell_1 \ell_2  \dots \ell_k$, the reward sequence equals $\mathcal{R}(\ell_1 \ell_2 \dots \ell_k)$.

Given a labeled MDP $\mathcal{M}=(S, s_I, A, R, P, \gamma, \mathcal{P}, L)$ with a non-Markovian reward function and a reward machine $\mathcal{R}=(U, u_I, \mathcal{P}, \delta, \sigma)$ which encodes the reward function $R$ of $\mathcal{M}$, we can obtain a product MDP $\mathcal{M}_{\mathcal{R}}$,  whose reward function is Markovian such that every attainable label sequence of $\mathcal{M}_{\mathcal{R}}$ gets the same reward as in $\mathcal{M}$. Furthermore, any policy for $\mathcal{M}_{\mathcal{R}}$ achieves the same expected reward in $\mathcal{M}$\cite{xu2020joint}. We define the product MDP $\mathcal{M}_R = (S', s_I', A, R', P', \gamma', \mathcal{P}', L')$ by:
\begin{equation}
    \begin{aligned}
        S' &= S \times U \\
        s_I' &= (s_I, u_I) \\
        A &= A \\
        R'(s, u, a) &=  \sigma(u, L(s, a, s'))\\
        P'( s, u, a, s', u') &= \begin{cases}
                                        P(s, a, s')  & u' = \delta(u, L(s, a, s'))\\
                                        0  & \mbox{otherwise}
                                    \end{cases} \\
        \gamma' &= \gamma; \mathcal{P}' = \mathcal{P}; L' = L
    \end{aligned}
    \label{product_Markovian}
\end{equation}

\subsection{Graph-Based Multi-Agent MDP with Reward Machines}

We denote $G = (V, E)$ as an undirected graph, where $V =\{1, \dots, n \}$ is a finite set of nodes that represents the agents and $E$ is a finite set of edges. For agents $i, j$, we call $j$ a neighboring agent of $i$, if there is an edge that connects $i$ and $j$. Let $N(i) \subseteq V$ be the set including agent $i$ and the neighboring agents of the agent $i$.

We consider a multi-agent system, including $n$ agents and formulate the system as a labeled graph-based MDP $\mathcal{M}_G = (S_G, s_{I_G}, A_G, R_G, P_G, \gamma, \mathcal{P}_G, L_G) = (\{S_i\}_{i=1}^n, \{s_I^i\}_{i=1}^n,$ $ \{A_i\}_{i=1}^n, \{R_i\}_{i=1}^n, \{P_i\}_{i=1}^n, \gamma, \{\mathcal{P}_i\}_{i=1}^n,
\{L_i\}_{i=1}^n)$. We consider a factored form for $\mathcal{M}_G$, i.e, $S_G$ is a Cartesian product of the states for each agent $i$ in $V$, $S_G = S_1 \times \dots \times S_n$. Similarly, $s_{I_G} = s_I^1 \times \dots \times s_I^n$; $A_G = A_1 \times \cdots \times A_n$; $R_G = R_1 \times \cdots \times R_n$; $P_G = P_1 \times \cdots \times P_n$; $\mathcal{P}_G = \mathcal{P}_1 \times \cdots \times \mathcal{P}_n$; and $L_G = L_1 \times \cdots \times L_n$. For each agent $i$, $P_i: S_{N(i)} \times A_{N(i)} \times S_i \to [0, 1]$, meaning $s_i(t+1)$ depends on $s_{N(i)}(t) \in S_{N(i)}$ and $a_{N(i)}(t) \in A_{N(i)}$, where $S_{N(i)}$ and $A_{N(i)}$ denote the Cartesian product of the sets of states and actions of the agents in $N(i)$ respectively. $R_i$ and $L_i$ only depend on the information of agent $i$ itself, $R_i: S_i \times A_i \to \mathbb{R}$ and $L_i: S_i \times A_i \times S_i \to 2^{\mathcal{P}_i}$. 

We design a reward machine $\mathcal{R}_i$ for each agent $i$. $\mathcal{R}_i = (U_i, u_I^i, \mathcal{P}_i, \delta_i, \sigma_i)$ encodes the reward function $R_i$ of $\mathcal{M}_G$ individually. $\mathcal{R}_i$ defines the task of agent $i$ in terms of high-level events from $\mathcal{P}_i$. Given a labeled graph-based MDP $\mathcal{M}_G=(S_G, s_{I_G}, A_G, R_G, P_G, \gamma, \mathcal{P}_G, L_G)$ with a non-Markovian reward function collectively defined by Reward Machines $\mathcal{R}_i$, we can obtain a product MDP $\mathcal{M}_{G \mathcal{R}} = (S_{G \mathcal{R}}, s_{I_{G \mathcal{R}}}, A_{G \mathcal{R}}, R_{G \mathcal{R}}, P_{G \mathcal{R}}, \gamma_{G \mathcal{R}}, \mathcal{P}_{G \mathcal{R}}, L_{G \mathcal{R}})$ whose reward function is Markovian, 
\begin{equation}
    \begin{aligned}
        S_{G \mathcal{R}} &= S_G \times U_1 \times \dots \times U_n \\
        s_{I_{G \mathcal{R}}} &= (s_{I_G}, u_I^1 \times \dots \times u_I^n) \\
        A_{G \mathcal{R}} &= A_G \\
        R_{G \mathcal{R}_i}(s_i, u_i, a_i) &=  \sigma_i(u_i, L_i(s_i, a_i, s'_i))\\
        P_{G \mathcal{R}}(s, u, a, s', u') &=  P_{G \mathcal{R}_1} \times \dots \times P_{G \mathcal{R}_n}\\
        P_{G \mathcal{R}_i}(s_{N(i)}, u_{N(i)}, a_{N(i)}, s_i', u_i')
        &= \begin{cases}
                                        P_i(s_{N(i)}, a_{N(i)}, s_i')  & \text{if} \quad u_i' = \delta(u_i, L_i(s_i, a_i, s_i'))\\
                                        0  & \mbox{otherwise}
                                    \end{cases} \\
        \gamma_{G \mathcal{R}} &= \gamma; \mathcal{P}_{G \mathcal{R}} = \mathcal{P}_G; L_{G \mathcal{R}} = L_G
    \end{aligned}
    \label{multi_product_Markovian}
\end{equation}
For simplicity of notation, we define the global state $s = (s_1, \dots, s_n) \in S_G$, global RM state $u = (u_1, \dots, u_n) \in U_G:= U_1 \times \dots \times U_n$, global action $a = (a_1, \dots, a_n) \in A_G$, and global reward $R(s, a) = \frac{1}{n} \sum_{i=1}^{n} R_i(s_i, a_i)$.

$\mathcal{M}_{G \mathcal{R}}$ augments RM state and MDP state, which tracks the progress through the task for each agent. The goal in an $\mathcal{M}_{G \mathcal{R}}$ is to find the optimal joint policy $\pi^\theta: S_{G \mathcal{R}} \to Distr(A_{G \mathcal{R}})$, which is parameterized by $\theta$. $\pi^\theta(a|s, u)$ is expected to maximize the discounted global reward, $J(\theta)$ starting from the initial state distribution $s_I$ and $u_I$,
\begin{equation}
    \max_\theta J(\theta) = \mathbb{E}_{(s, u) \sim s_I, u_I} \mathbb{E}_{a(t) \sim \pi^\theta (\cdot |s(t), u(t))} [\sum_{t=0}^{\infty} \gamma^t R(s(t), u(t), a(t) | s(0)  = s_I, u(0) = u_I]
\end{equation}
The Q-function under the policy $\pi^\theta$, $Q^\theta(s, u, a)$, is defined as the expected discounted future
reward of taking action $a$ given a pair $(s, u)$ and then following the policy $\pi^\theta$,
\begin{equation}
    \begin{aligned}
        Q^\theta(s, u, a) 
        &= \mathbb{E}_{a(t) \sim \pi^\theta (\cdot |s(t), u(t))} [\sum_{t=0}^{\infty} \gamma^t
         R(s(t), u(t), a(t) |
         s(0)  = s_I, u(0) = u_I, a(0) =a] \\
        &= \frac{1}{n} \sum_{i=1}^n \mathbb{E}_{a(t) \sim \pi^\theta (\cdot |s(t), u(t))}
        [\sum_{t=0}^{\infty} \gamma^t R_i (s_i(t), u_i(t), a_i (t) | 
         s(0) = s, u(0) = u, a(0) =a ] \\
        & =  \frac{1}{n} \sum_{i=1}^n Q_i^\theta(s,u, a)
    \end{aligned}
\end{equation}
where $Q_i^\theta(s,u, a)$ is the Q-function for the individual reward $R_i$. Compared to standard Q-learning, $Q^\theta(s, u, a)$ also keeps track of RM states, allowing agents to take the high-level stages into consideration when selecting actions. 

A commonly used RL algorithm is policy gradient \cite{sutton2000policy}. Let $d^\theta$ be a distribution on the $S_{G \mathcal{R}}$ given by $d^\theta(s, u) = (1-\gamma) \sum_{t=0}^\infty \gamma^t d_t^\theta(s, u)$, where $d_t^\theta(s, u)$ is the distribution of $(s(t), u(t))$ under fixed policy $\theta$ when $(s(0), u(0))$ is drawn from $d_0$. Then,
\begin{equation}
        \nabla J(\theta) = \frac{1}{1-\gamma} \mathbb{E}_{(s,u) \sim d^\theta, a \sim \pi^\theta(\cdot |s,u)} Q^\theta(s,u,a)\nabla \log \pi^\theta(a|s,u)
\end{equation}
Since the sizes of the joint state and action spaces, $S_{G \mathcal{R}}$ and $A_{G \mathcal{R}}$, grow exponentially with the number of agents, the corresponding Q-functions $Q^\theta$ and $Q_i^\theta$ can be large tables and therefore, are intractable to compute and store.   

\section{Decentralized Graph-Based Reinforcement Learning Using Reward Machines (DGRM)}

We propose the truncated Q-function ($\Tilde{Q}_i^\theta$), which takes local information of each agent as the input, and thus, the dimension of $\Tilde{Q}_i^\theta$ is much smaller than $Q_i^\theta$. We present that $\Tilde{Q}_i^\theta$ is able to approximate $Q_i^\theta$ with high accuracy. Then, we propose the truncated policy gradient based on $\Tilde{Q}_i^\theta$ and the error of policy gradient approximation is bounded. Moreover, the \foo ABC algorithm can find an $O(\rho^{\kappa+1})$-approximation of a stationary point of $J(\theta)$. To further improve efficiency, we also develop the deep \foo ABC algorithm, using deep neural networks to approximate the Q-function and policy function to solve large-scale or continuous state problems.


\subsection{\foo ABC algorithm}

To reduce complexity, we propose a decentralized learning algorithm that equips each agent a localized policy $\pi_i^{\theta_i}$, parameterized by $\theta_i$. The localized policy for agent $i$,  $\pi_i^{\theta_i} (a_i | s_i, u_i)$, is a distribution on the local action $a_i$, depending on the local state $s_i$ and local RM state $u_i$. Thus, each agent acts independently and information exchange is not required with learned policies. Moreover, the RM state is augmented with low-level MDP state for the policy, which helps provide guidance to the agent in the task level. The joint policy, $\pi^\theta(a|s, u)$, is defined as $\prod_{i=1}^n \pi_i^{\theta_i} (a_i|s_i, u_i), \theta = (\theta_1, \dots,\theta_n)$. $\pi_i^{\theta_i}$ is similar to the policy defined in the Scalable Actor Critic (SAC) algorithm
\cite{qu2020scalable}, which is conditioned only on the local state $s_i$. $\pi(a|s, u)$ differs from the policy in graph-based MDP \cite{cheng2013variational} in that their policy is conditioned on the agent's neighborhood. 

Further, \foo ABC algorithm uses local information of each agent $i$ to approximate $Q_i^\theta$, inspired by \cite{qu2020scalable}. We assume that the agent $i$ has access to the neighborhood of radius $\kappa$ (or $\kappa$-hop neighborhood) of agent $i$, denoted by $N_i^{\kappa}$, $\kappa \ge 0$. $\kappa$-hop neighborhood of $i$ includes the agents whose shortest graph distance to agent $i$ is less than or equal to $\kappa$, including agent $i$ itself. We define $N_{-i}^\kappa=V/N_i^k$, meaning the set of agents that are outside of agent $i$'s $\kappa$-hop neighborhood. The global state $s$ can also be written as $(s_{N_i^\kappa}, s_{N_{-i}^\kappa})$, which are the states of agents that are inside and outside agent $i$'s $\kappa$-hop neighborhood respectively, similarly, the global RM state $u = (u_{N_i^\kappa}, u_{N_{-i}^\kappa})$ and the global action $a=(a_{N_i^\kappa}, a_{N_{-i}^\kappa})$. We first define the exponential decay property of Q-function.

\textbf{Definition 2.} Q-function has the $(\lambda, \rho)$-exponential decay property if, for any localized policy $\theta$, for any $i \in V, s_{N_i^\kappa} \in S_{G_{N_i^\kappa}}, s_{N_{-i}^\kappa}, s_{N_{-i}^\kappa}' \in S_{G_{N_{-i}^\kappa}}, u_{N_i^\kappa} \in U_{G_{N_i^\kappa}}, u_{N_{-i}^\kappa}, u_{N_{-i}^\kappa}' \in U_{G_{N_{-i}^\kappa}}, a_{N_i^\kappa} \in A_{G_{N_i^\kappa}}, a_{N_{-i}^\kappa}, a_{N_{-i}^\kappa}' \in A_{G_{N_{-i}^\kappa}}$, $Q_i^\theta$ satisfies,
\begin{equation}
    |Q_i^\theta(s_{N_i^\kappa}, s_{N_{-i}^\kappa}, u_{N_i^\kappa}, u_{N_{-i}^\kappa}, a_{N_i^\kappa}, a_{N_{-i}^\kappa})   - Q_i^\theta(s_{N_i^\kappa}, s_{N_{-i}^\kappa}', u_{N_i^\kappa}, u_{N_{-i}^\kappa}', a_{N_i^\kappa}, a_{N_{-i}^\kappa}')| \le \lambda \rho^{\kappa + 1}
\label{eq:decay0}
\end{equation}

\textbf{Theorem 1.} Assume $\forall i$, $R_i$ is upper bounded by $\overline{R}$, the $(\frac{\overline{R}}{1-\gamma}, \gamma)$-exponential decay property of $Q_i^\theta$ holds. 

The proof of Theorem 1 can be found in the supplementary material. Theorem 1 indicates that the dependency of $Q_i^\theta$ on other agents decreases exponentially as the distance between them grows. Then, we propose the following truncated Q-functions for state-(RM state)-action triples,
\begin{equation}
\begin{aligned}
    \Tilde{Q}_i^\theta(s_{N_i^\kappa}, u_{N_i^\kappa}, a_{N_i^\kappa}) &= \sum_{s_{N_{-i}^\kappa} \in S_{G_{N_{-i}^\kappa}}, u_{N_{-i}^\kappa} \in U_{G_{N_{-i}^\kappa}}, a_{N_{-i}^\kappa} \in A_{G_{N_{-i}^\kappa}}}  c_i(s_{N_{-i}^\kappa},u_{N_{-i}^\kappa},a_{N_{-i}^\kappa}; s_{N_i^\kappa},u_{N_i^\kappa},a_{N_i^\kappa}) \\ 
    & \quad \cdot Q_i^\theta(s_{N_i^\kappa},s_{N_{-i}^\kappa},u_{N_i^\kappa},u_{N_{-i}^\kappa},a_{N_i^\kappa},a_{N_{-i}^\kappa})
\end{aligned}
\label{eq:truncated_Q}
\end{equation}
where $c_i(s_{N_{-i}^\kappa},u_{N_{-i}^\kappa},a_{N_{-i}^\kappa}; s_{N_i^\kappa},u_{N_i^\kappa},a_{N_i^\kappa})$ is any non-negative weight and satisfies,

\begin{equation}
\begin{aligned}
   & \forall (s_{N_i^\kappa},u_{N_i^\kappa},a_{N_i^\kappa}) \in S_{G_{N_i^\kappa}} \times U_{G_{N_i^\kappa}} \times A_{G_{N_i^\kappa}}, \\
   &\sum_{s_{N_{-i}^\kappa} \in S_{G_{N_{-i}^\kappa}}, u_{N_{-i}^\kappa} \in U_{G_{N_{-i}^\kappa}}, a_{N_{-i}^\kappa} \in A_{G_{N_{-i}^\kappa}}}  c_i(s_{N_{-i}^\kappa},u_{N_{-i}^\kappa},a_{N_{-i}^\kappa}; s_{N_i^\kappa},u_{N_i^\kappa},a_{N_i^\kappa}) =1.
\end{aligned}
\end{equation}

\textbf{Theorem 2.} Under the $(\lambda, \rho)$-exponential decay property, $\Tilde{Q}_i^\theta$ satisfies,
\begin{equation}
    \sup_{(s, u, a) \in S_G \times U_G \times A_G} |\Tilde{Q}_i^\theta(s_{N_i^\kappa}, u_{N_i^\kappa}, a_{N_i^\kappa})-Q_i^\theta(s, u,a)| \le \lambda \rho^{\kappa + 1}
\end{equation}

The proof of Theorem 2 can be found in the supplementary material. Theorem 2 shows that $\Tilde{Q}_i^\theta$ approximates $Q_i^\theta$ with high accuracy, and next, we use $\Tilde{Q}_i^\theta$ to approximate the policy gradient. The truncated policy gradient for agent $i$, $\Tilde{J}_i$, is defined as $\Tilde{J}_i(\theta) = \frac{1}{1-\gamma} \mathbb{E}_{(s,u) \sim d^\theta, a \sim \pi^\theta(\cdot |s,u)} [\frac{1}{n} \sum_{j \in N_i^\kappa} \Tilde{Q}_j^\theta(s_{N_j^\kappa}, u_{N_j^\kappa}, a_{N_j^\kappa})] \nabla_{\theta_i} \log \pi_i^{\theta_i}(a_i|s_i. u_i)$, where $\Tilde{Q}_j^\theta$ is any truncated Q-function in the form of \cref{eq:truncated_Q}.

\textbf{Theorem 3.} If $||\nabla_{\theta_i}\log \pi_i^{\theta_i}(a_i|s_i, u_i)||$ is bounded by $B_i$, $\forall a_i, s_i, u_i$, $||\Tilde{J}_i(\theta) - \nabla_{\theta_i} J(\theta)|| \le \frac{\lambda B_i}{1-\gamma} \rho^{\kappa+1}$

The proof of Theorem 3 can be found in the supplementary material. Theorem 3 indicates that the error of the truncated policy gradient is bounded and thus, it is safe to use the truncated Q-function to approximate the policy gradient, which has a much smaller dimension.

The \foo ABC algorithm adopts the actor-critic framework, specifically using temporal difference (TD) learning to train the truncated Q-function and policy gradient for the localized policy. The pseudocode of \foo ABC is given in \cref{alg:alg1}. The truncated Q-function is updated at every time step, and the policy parameters are updated in every episode. \foo ABC reduces the computational complexity as the truncated Q-function is conditioned on the state-(RM state)-action triple of $\kappa$-hop neighborhood instead of the global information, which results in a much smaller dimension. Thus, the complexity of \foo ABC is the local state-(RM state)-action space size of the largest $\kappa$-hop neighborhood. The approximated gradient of policy parameters is related to the truncated Q-function of the $\kappa$-hop neighborhood. Further, the authors in \cite{qu2020scalable} claimed that under certain assumptions, the actor-critic algorithm will eventually find an $O(\rho^{\kappa+1})$-approximation of a stationary point of $J(\theta)$. \foo ABC also follows the assumptions with RM state, thus the approximating convergence conclusion still holds, which is illustrated in Theorem 4.


\textbf{Assumption 1.} The reward is bounded, $R_i(s_i,u_i, a_i) \le \overline{R}, \forall i, s_i, u_i, a_i$. The individual state and action space size are upper bounded as $|S_i| \le \overline{S}, |A_i | \le \overline{A}, \forall i$.

\textbf{Assumption 2.} The $(\lambda, \rho)$ exponential decay property holds for some $\rho \le \gamma$.

\textbf{Assumption 3.} Every $(s_{N_i^\kappa}, u_{N_i^\kappa}, a_{N_i^\kappa})$ pair must be visited with some positive probability after some time.

\textbf{Assumption 4.} For any $i, a_i, s_i, u_i$ and $\theta_i$, we assume $||\nabla_{\theta_i}\log \pi_i^{\theta_i}(a_i|s_i, u_i)|| \le B_i$. As a result, $||\nabla_{\theta}\log \pi^{\theta}(a|s, u)|| \le B = \sqrt{\sum_{i=1}^n B_i^2}$. Further, assume $\nabla J(\theta)$ is $B'$-Lipschitz continuous in $\theta$.

\textbf{Theorem 4.} Under Assumption 1, 2, 3, and 4, \foo ABC will find an $O(\rho^{\kappa+1})$-approximation of a stationary point of the objective function $J(\theta)$.

\begin{algorithm}[tb]
\textbf{Input}: initial policy $\pi_i^{\theta_i}(0)$, initial Q-function, $s(0)$, $u(0)$, $\kappa, T, \gamma, \alpha_{\Tilde{Q}}, \alpha_\pi$.
\begin{algorithmic}[1]
\FOR{each Episode $\Tilde{e}$}
 \STATE
 Take action $a_i(0) \sim \pi_i^{\theta_i}(\cdot | s_i(0), u_i(0))$\ for all $i$.\\
 \STATE Get reward $R_i(0)$ for all $i$.\\
 \WHILE{exist agent not finish the task and $t \le T$} 
  \STATE Get state $s_i(t)$, RM state $u_i(t)$ for all $i$.\\
  \STATE Take action $a_i(t) \sim \pi_i^{\theta_i}(\cdot | s_i(t), u_i(t))$\ for all $i$.\\
  \STATE Get reward $R_i(t)$ for all $i$.\\
  \STATE Calculate TD error for all $i$,\\
  \STATE $TD_i \gets R_i(t-1) + \gamma \Tilde{Q}_i^{t-1} \big( s_{N_i^k}(t), u_{N_i^k}(t),  a_{N_i^k} (t)\big)- \Tilde{Q}_i^{t-1}\big(s_{N_i^k}(t-1), u_{N_i^k}(t-1), a_{N_i^k}(t-1)\big)$. \\
  \STATE Update the truncated Q-function for all $i$,\\
  \STATE $\Tilde{Q}_i^{t}\big(s_{N_i^k}(t-1), u_{N_i^k}(t-1), a_{N_i^k}(t-1)\big) \gets \Tilde{Q}_i^{t-1}\big(s_{N_i^k}(t-1), u_{N_i^k}(t-1), a_{N_i^k}(t-1)\big) + \alpha_{{\Tilde{Q}}_{t-1}} TD_i$.
  \ENDWHILE
  \STATE Calculate the approximated gradient for all $i$,\\
  \STATE $\Tilde{g}_i(\Tilde{e}) \gets  \sum_{t'=0}^t \gamma^t \frac{1}{n} \sum_{j \in N_i^k} \Tilde{Q}_j^t \big (s_{N_i^k}(t'), u_{N_i^k}(t'), a_{N_i^k}(t')\big) \nabla_{\theta_i} \log  \pi_i^{\theta_i(\Tilde{e})}\big(a_i(t')|s_i(t'), u_i(t')\big)$.\\
  \STATE Update the policy parameters for all agents,\\
  \STATE $\theta_i(\Tilde{e}+1) \gets \theta_i(\Tilde{e}) + \alpha_{\pi_{\Tilde{e}}} \Tilde{g}_i(\Tilde{e}) $.
\ENDFOR
\end{algorithmic}
\caption{\foo ABC Algorithm}
\label{alg:alg1}
\end{algorithm}

\subsection{Deep \foo ABC algorithm}

To further improve efficiency, we approximate both the policy and truncated Q-functions by deep neural networks. The pseudocode of deep \foo ABC is given in \cref{alg:alg2}. In every episode from line 2 to line 7, all the agents first interact with the environment. Then in line 8, each agent collects the necessary information during the episode for later actor and critic updating. Both the Q-network and policy-network are updated in every episode. In line 11, the TD error is the averaged error over one episode. In line 15, the gradient of policy is conditioned on the averaged TD error of the $\kappa$-hop neighborhood, which indicates communication is required for policy improvement during training. When implementing the learned policy after training, communication is not required since the policy is only conditioned on each agent's own state-(RM state) pair.

\begin{algorithm}[tb]
\textbf{Input:} initialized policy $\theta_i(0)$, initialized Q-network $\beta_i(0)$, $s(0)$, $u(0)$, $\kappa$, $T$, $\gamma$, $\alpha_{\Tilde{Q}}$, $\alpha_\pi$ 
\begin{algorithmic}[1]
\FOR{each Episode $\Tilde{e}$}
 \STATE Take action $a_i(0) \sim \pi_i^{\theta_i}(\cdot | s_i(0), u_i(0))$\ for all $i$.\\
 \STATE Get reward $R_i(0)$ for all $i$.\\
 \WHILE{exist agent not finish the task and $t \le T$}
 \STATE Get state $s_i(t)$, RM state $u_i(t)$ for all $i$.\\
 \STATE Take action $a_i(t) \sim \pi_i^{\theta_i}(\cdot | s_i(t), u_i(t))$\ for all $i$.\\
 \STATE Get reward $R_i(t)$ for all $i$.\\
 \STATE Store transition $\big(s_{N_i^\kappa}(t-1), u_{N_i^\kappa}(t-1), a_{N_i^\kappa}(t-1), R_i(t-1), s_{N_i^\kappa}(t), u_{N_i^\kappa}(t), a_{N_i^\kappa}(t)\big)$ for all $i$.
 \ENDWHILE
 \STATE Calculate the TD error for all $i$, \\
 \STATE $TD_i \gets \frac{1}{t} \sum_{t'=1}^t \Big(R_i(t'-1) + \gamma \Tilde{Q}_i^{\Tilde{e}}\big(s_{N_i^\kappa}(t'), u_{N_i^\kappa}(t'),
 a_{N_i^\kappa}(t')\big) - \Tilde{Q}_i^{\Tilde{e}}\big(s_{N_i^\kappa}(t'-1), u_{N_i^\kappa}(t'-1), a_{N_i^\kappa}(t'-1)\big)\Big)$. \\
 \STATE Update Q-network parameters for all $i$,\\
 \STATE $\beta_i(\Tilde{e}+1) \gets \beta_i(\Tilde{e}) + \alpha_{\Tilde{Q}} TD_i \nabla_{\beta_i} \Tilde{Q}_i^{\Tilde{e}}$.\\
 \STATE Update the policy-network parameters for all $i$,\\
 \STATE $\theta_i(\Tilde{e}+1) \gets \theta_i(\Tilde{e}) + \alpha_{\pi} \frac{1}{|N_i^{\kappa}|} \sum_{j \in N_i^{\kappa}} TD_j \nabla_{\theta_i} \ln \pi_i$.
\ENDFOR
\end{algorithmic}
\caption{Deep \foo ABC Algorithm}
\label{alg:alg2}
\end{algorithm}

\section{Case Studies}

In this section, we implement \foo ABC on two case studies: (1) a multi-UAV package delivery case and (2) COVID-19 pandemic mitigation adapted from \cite{della2020intermittent}.

\subsection{Case Study I: Multi-UAV Package Delivery}

We consider a total of six UAVs in the 4$\times$5 grid-world with two warehouses. The task for each agent is to obtain one package in a warehouse and deliver it to the corresponding destination. Further details of the grid-world environment are illustrated in the supplementary material.

The local state of UAV $i$ at time $t$, $s_i(t)$, includes UAV $i$'s current position, the remaining battery percentage and a binary variable with value of 1 indicating UAV $i$ obtains a package. UAVs start with a fully charged battery. 
UAVs have four base possible actions at all grids: move north, move south, move east, and move west. At warehouses, UAVs have two additional possible actions: wait and pick up. When waiting, the agent remains where it was. The agent obtains the package with a probability of 0.9 if it selects to pick up and no other agents also pick up at the same warehouse. We define two UAVs are neighbors if they have access to the same warehouse.

\begin{figure}[hbt!]
    \centering
    \subfigure[]{\includegraphics[width=0.45\textwidth]{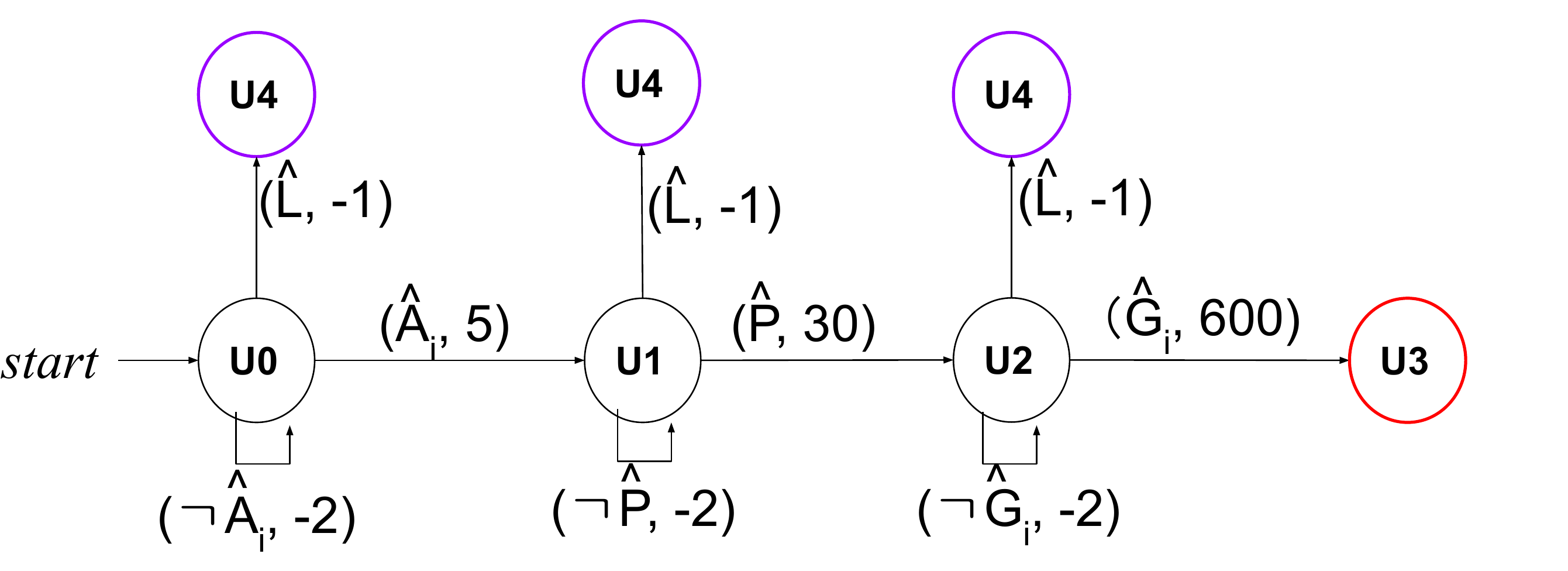}\label{RM_grid_1}} 
    \hspace{1cm}
    \subfigure[]{\includegraphics[width=0.45\textwidth]{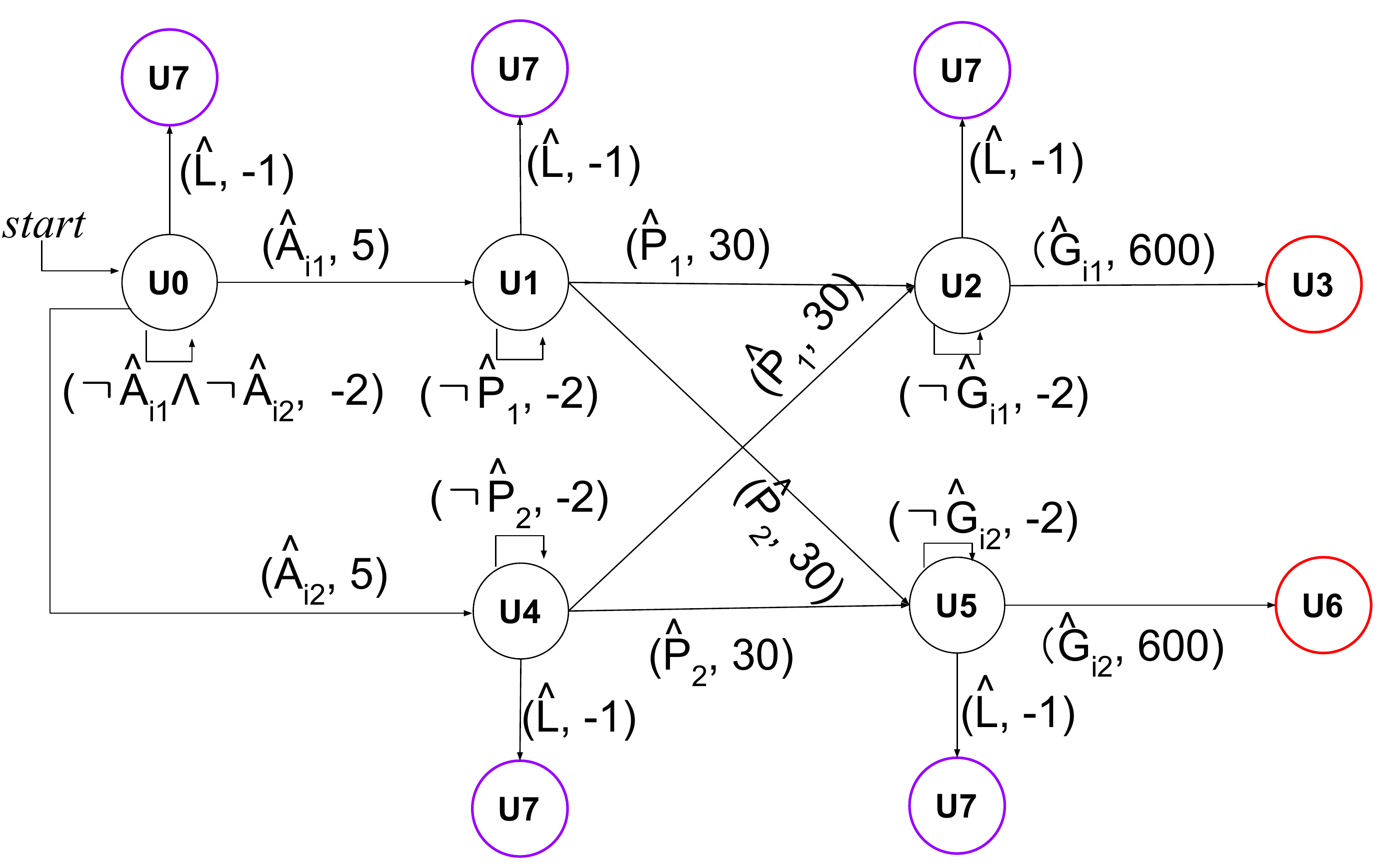}\label{RM_grid_2}} 
    \caption{(a) Reward machine for UAV $i\in \{1,2,5,6 \}$ in case study I. (b) Reward machine for UAV $i\in \{3, 4 \}$ in case study I. Purple: sink state; Red: goal state.}
    \label{RM_grid}
\end{figure}

UAV $i\in{1, 2, 5, 6}$ has access to one specific warehouse and UAV $i\in{3, 4}$ has access to two warehouses. The reward machine for UAV $i$ is shown in \cref{RM_grid}. Each edge is labeled by a tuple $(\hat{p}, \hat{r})$, where $\hat{p}$ is a propositional logic formula over $\mathcal{P}$ and $\hat{r} \in \mathbb{R}$ is a real number, defining the reward of the corresponding transition. For instance, to transition from $u_0$ to $u_1$ in \cref{RM_grid_1}, the proposition $\hat{A}$ has to be true, and the transition outputs a reward of 5. Regarding UAV $i\in \{1,2,5,6 \}$, the labeled events are $\mathcal{P}_i = \{\hat{A_i}, \hat{P}, \hat{G_i}, \hat{L} \}$, where $\hat{A_i}$ means UAV $i$ reaches the accessible warehouse, $\hat{P_i}$ indicates that UAV $i$ obtains the package, $\hat{G_i}$ is arriving at the destination, and $\hat{L}$ indicates that UAV is running out of battery. When the agent has low battery, it transitions to the sink state $u_4$. Once the agent reaches a sink state, it cannot get out of the sink state and fails the task. $u_3$ is the goal state, expressing task accomplishment. Since UAV $i\in \{3, 4\}$ has access to two warehouses, UAV 3 and UAV 4 have two possible routes to accomplish the task. Further details of the reward machine of UAV 3 and UAV 4 are illustrated in the supplementary material. 

\begin{figure}[t]
\centering
\includegraphics[width=0.6\textwidth]{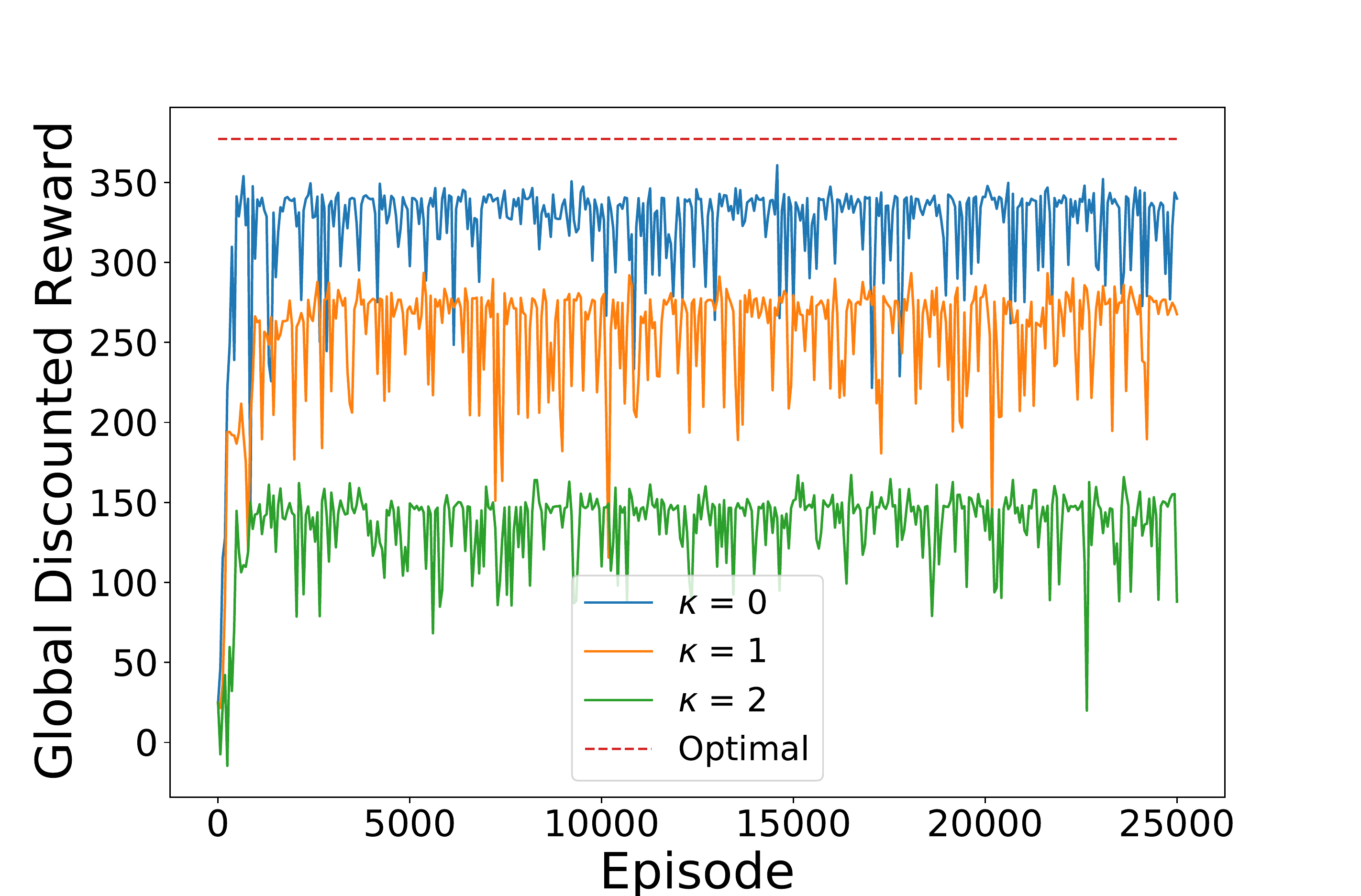}
\caption{Global discounted rewards during the training process in case study I (denoised using B-spline interpolation).}
\label{grid_k}
\end{figure}

In the experiment, we set $\gamma=0.9$. The softmax policy is utilized for the localized policy, which forms parameterized numerical probabilities for each state–action pair \cite{sutton2018reinforcement}. We run the DRGM algorithm with $\kappa \in \{0, 1, 2\}$ and plot the global discounted reward during the training process in \cref{grid_k}. The optimal solution is in closed form for this experiment. \foo ABC with $\kappa = 0$ has the best performance and is close to the optimal solution. Also, \foo ABC with $\kappa = 0$ converges the fastest. \cref{grid_k} indicates that more information of neighboring agents does not help improve the performance in this example. The possible reason is that we use the tabular Q-function for this discrete state problem and a larger $\kappa$-hop neighborhood introduces large state, RM state, and action space for the tabular Q-function, which may cause poor performance. Additionally, the result in \cref{grid_k} indicates that \foo ABC has the potential to determine the necessary information for the truncated Q-function and help understand the connections among the agents. 


\subsection{Case Study II: COVID-19 Pandemic Mitigation}
 
To the best of our knowledge, this case study provides the first attempt to use reward machines as specifications to mitigate epidemics and pandemics such as COVID-19.

We consider a network model of Italy \cite{della2020intermittent}, where Italy is modeled as a network of 20 regions and the parameters of each region's model are calculated from real data. The total population is divided into five subgroups for each region $i$: susceptible ($\mathcal{S}_i$), infected ($\mathcal{I}_i$), quarantined ($\mathcal{Q}_i$), hospitalized ($\mathcal{H}_i$), recovered ($\mathcal{C}_i$), and deceased ($\mathcal{D}_i$). The detailed analysis of real data and the estimation of parameters of the pandemic model can be referred to \cite{della2020intermittent}.


The authors in \cite{della2020intermittent} propose a bang-bang control strategy for each region according to the relative saturation level of its health system. The regional lockdown is enforced when the relative saturation level is high. During the lockdown, the region implements strict social distancing rules, and control all fluxes in or out of the region. Further details of the bang-bang control and mathematical derivations are illustrated in the supplementary material. We take the above bang-bang control as a baseline method for comparison with our proposed method.

In this case study, we aim to control the spread of COVID-19 among the 20 regions of Italy in a period of 4 weeks. A region is defined to be in a severe situation when the total hospital capacity is saturated, denoted by $\frac{0.1\mathcal{H}_i}{\mathcal{T}_i^\mathcal{H}} \ge 0.5$, which is the same as the saturation level when a lockdown is adopted in the bang-bang control. $0.1\mathcal{H}_i$ estimates the number of the hospitalized requiring care in ICU and $\mathcal{T}_i^\mathcal{H}$ is the number of available ICU beds in the region $i$. The control objective is that each region would not be in a severe situation for 2 consecutive weeks and avoid long time lockdown of 2 consecutive weeks, since costs of continuing severe restrictions could be great relative to likely benefits in lives saved \cite{miles2021stay}.

To conduct our experiment, we establish a labeled graph-based multi-agent MDP for the network model of Italy and a reward machine as task specification for each region. We define that region $j$ is region $i$'s neighbor if there are fluxes of people traveling between two regions. The simulation is conducted with a discretization step of 1 day and starts on a Monday. At the first step pf the simulation, each region is set to be in a severe situation,  $\mathcal{H}_i(0) = 6\mathcal{T}_i^\mathcal{H}$,  the values of $\mathcal{I}_i(0)$, $\mathcal{Q}_i(0)$, $\mathcal{C}_i(0)$, and $\mathcal{D}_i(0)$ are the same as the experiment in \cite{della2020intermittent}. Each region has four possible actions at each time step: no restrictions, implementing social distancing, controlling fluxes between neighboring regions, and adopting a lockdown. The local state of region $i$ at day $t$, $s_i(t)$, includes the characterization of its population, high-level features of this week's situation, and the simulated time step, \begin{equation}
    s_i(t) = (\mathcal{S}_i(t), \mathcal{I}_i(t), \mathcal{R}_i(t), \mathcal{H}_i(t), \mathcal{Q}_i(t), \mathcal{D}_i(t), \Tilde{v}_i, \Tilde{l}_i, t)
\end{equation}
where $\Tilde{v}_i$ and $\Tilde{l}_i$ are the number of days when the region is in a severe situation and the number of days when the region adopts a lockdown since this Monday, respectively. The transitions of $\mathcal{S}_i(t), \mathcal{I}_i(t), \mathcal{R}_i(t), \mathcal{H}_i(t), \mathcal{Q}_i(t), \mathcal{D}_i(t)$ follow the pandemic model's dynamics in \cite{della2020intermittent}. The set of events is $\mathcal{P}_i=\{ \epsilon_0, \epsilon_1,  v^0l^0,  v^{0.5}l^0, v^1l^0, v^0l^{0.5},  v^{0.5}l^{0.5}, v^1l^{0.5}, v^0l^1,  v^{0.5}l^1, \\ v^1l^1\}$, where $\epsilon_0$ is an empty event when the current day is not Monday; $\epsilon_1$ is the event leading to the goal state and indicates the current day is the last time step of the entire simulation; the superscript of $v$ and $l$ describe the level of the situation's severity and the frequency of lockdown implementation during the previous seven days (from Monday to Sunday), respectively. Further details of $v$ and $l$ are illustrated in the supplementary material.

We note that the reward machine for each region is the same for simplicity. The initial RM state assumes that during the previous week, the region was in a severe situation at least 1 day but less than 7 days and did not implement a lockdown. We define sink states to indicate that the region has been in a severe situation or has implemented a lockdown for two consecutive weeks, which are not encouraged. Further details of the reward machine and the reward function are illustrated in the supplementary material.



\begin{figure}[H]
\centering
\includegraphics[width=0.5\textwidth]{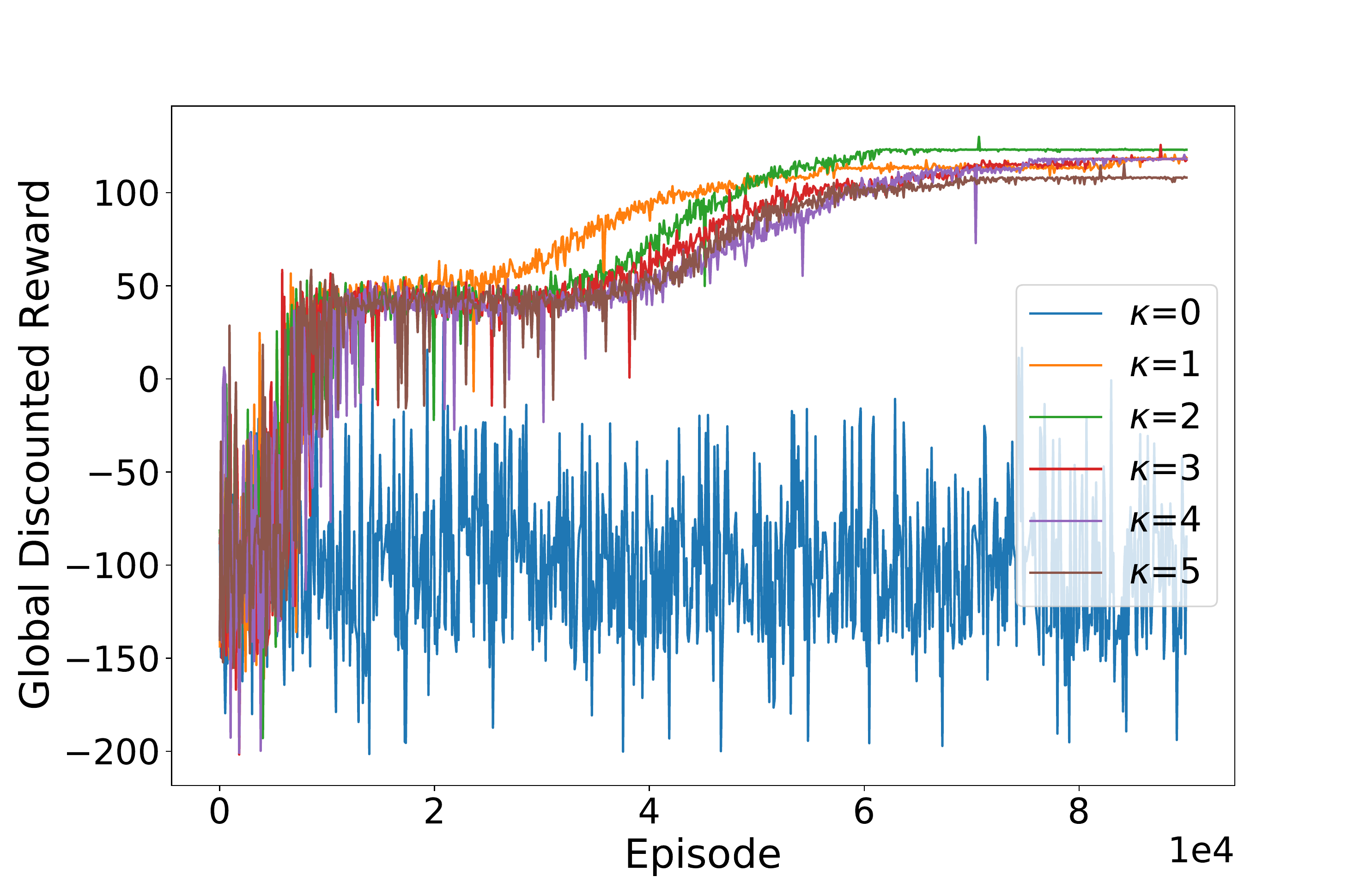}
\caption{Global discounted rewards during the training process in case study II (denoised using B-spline interpolation).}
\label{covid_k}
\end{figure}

\begin{figure}[H]
\centering
\includegraphics[width=0.5\textwidth]{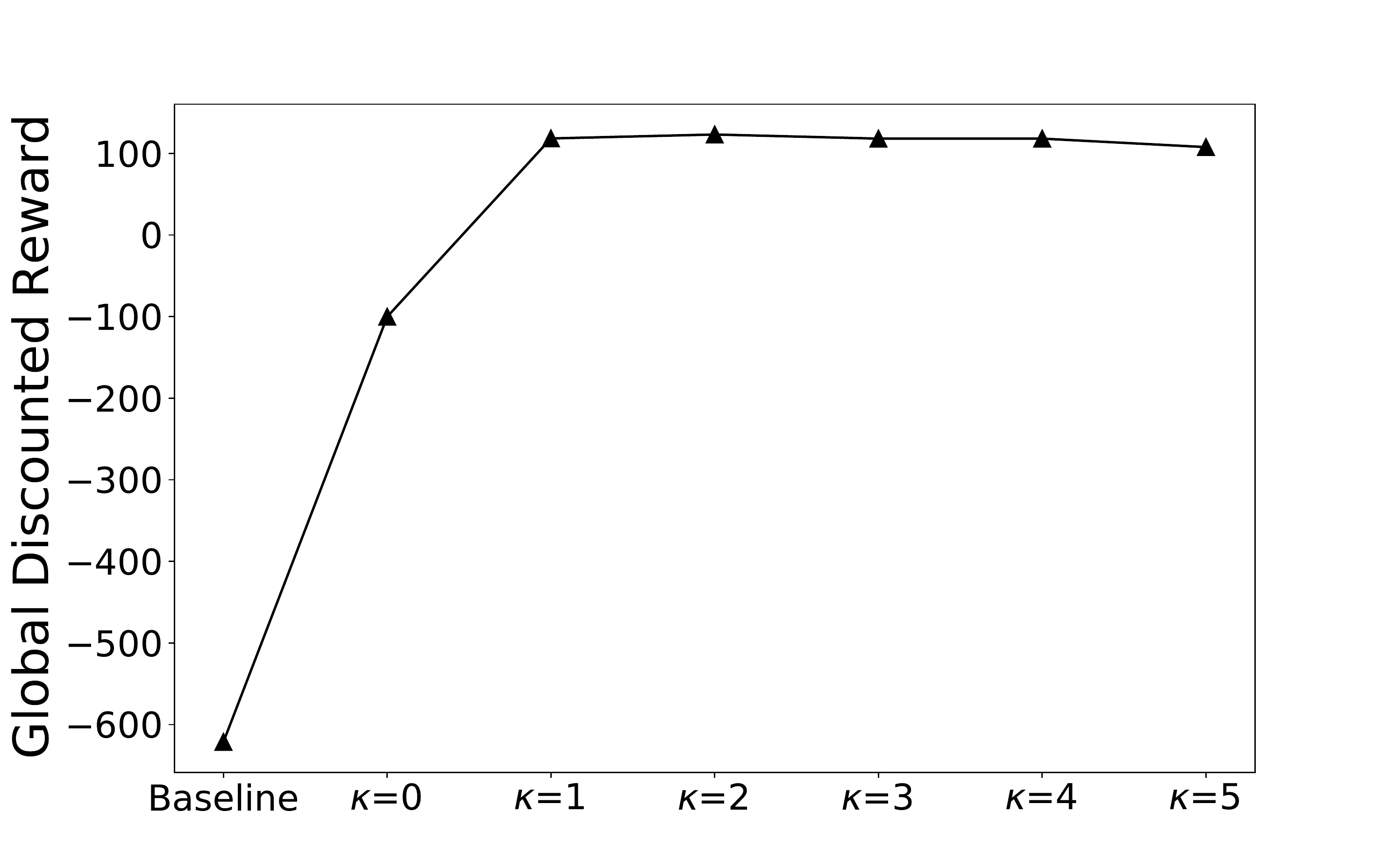}
\caption{Comparison of global discounted rewards in case study II.}.
\label{covid_r_bar}
\end{figure}

In the experiment, we set $\gamma = 0.9$. We use two fully connected layers with [256, 128] units and ReLU activations for actor and two fully connected layers with [256, 128] units and Tanh activation for critic. We run the deep \foo ABC algorithm with $\kappa = 0$ up to $\kappa = 5$ and plot the global discounted rewards during the training process in \cref{covid_k}. With the increase in $\kappa$, the global discounted reward increases when $\kappa \le 2$. This is also demonstrated in \cref{covid_r_bar}, which shows the performance of the baseline (bang-bang control policy in \cite{della2020intermittent}) and the results of deep \foo ABC obtained by 20 independent simulation runs using the converged policy for each $\kappa$. The deep \foo ABC algorithm with $\kappa > 0$ outperforms $\kappa = 0$, which is the independent learner method in \cite{tan1993multi}. The deep \foo ABC algorithm with $\kappa=1$ improves the global discounted reward by 218\% compared to $\kappa=0$. The deep \foo ABC with $\kappa=2$ has limited improvement (4\%) on the performance compared to $\kappa=1$. When $\kappa>2$, the global discounted reward stops growing due to the noise. This is also consistent with Theorem 4 that the proposed algorithm can reach a stationary point. We note that with the deep \foo ABC algorithm, information of 1-hop neighborhood is sufficient for this experiment and the global information is unnecessary for good performance. Moreover, the deep \foo ABC with $\kappa =1 $ improves the global accumulated reward by 119\% compared to the baseline. Further details of the result with $\kappa =1 $, including the local discounted rewards and the actions selected by each region are illustrated in the supplementary material.

\section{Conclusion}

In this work, reward machines are leveraged to express complex temporally extended tasks. A decentralized learning algorithm for graph-based MDP is introduced. During training, we use the local information of the $\kappa$-hop neighborhood for a truncated Q-function. The truncated Q-function can be computed efficiently, and we prove that the approximation error is bounded. After training, agents can implement the policies independently. We also combine the decentralized architecture and function approximation by deep neural networks to enable the application to large-scale MARL or continuous state problems. Moreover, this is the first work to use reward machines to mitigate the COVID-19 pandemic through task specification, and we show the performance improvement over the baseline method.

The current work assumes each agent has an individual reward machine and the reward machine is known, but it would be possible to learn reward machines from experience \cite{xu2020joint} and decompose a team-level task to a collection of reward machines \cite{neary2020reward}. We use a standard actor-critc algorithm, one possible research direction is to use more advanced RL algorithms such as Proximal Policy Optimization (PPO) \cite{schulman2017proximal} to allow for solving continuous action problems. Leveraging strategies such as Hindsight Experience Replay \cite{andrychowicz2017hindsight} to learn policies faster is also worth investigating.

\section*{Acknowledgments}
The research reported in this paper was supported by funds from NASA University Leadership Initiative program (Contract No. NNX17AJ86A, PI: Yongming Liu, Technical Officer: Anupa Bajwa). The support is gratefully acknowledged.

\bibliography{aaai22}

\begin{thebibliography}{40}
\newcommand{\enquote}[1]{``#1''}
\providecommand{\natexlab}[1]{#1}
\providecommand{\url}[1]{\texttt{#1}}
\providecommand{\urlprefix}{URL }
\expandafter\ifx\csname urlstyle\endcsname\relax
  \providecommand{\doi}[1]{doi:\discretionary{}{}{}#1}\else
  \providecommand{\doi}{doi:\discretionary{}{}{}\begingroup
  \urlstyle{rm}\Url}\fi

\bibitem[{Dean and Kanazawa(1989)}]{dean1989model}
Dean, T., and Kanazawa, K., \enquote{A model for reasoning about persistence
  and causation,} \emph{Computational intelligence}, Vol.~5, No.~2, 1989, pp.
  142--150.

\bibitem[{Guestrin et~al.(2003)Guestrin, Koller, Parr, and
  Venkataraman}]{guestrin2003efficient}
Guestrin, C., Koller, D., Parr, R., and Venkataraman, S., \enquote{Efficient
  solution algorithms for factored MDPs,} \emph{Journal of Artificial
  Intelligence Research}, Vol.~19, 2003, pp. 399--468.

\bibitem[{Forsell and Sabbadin(2006)}]{forsell2006approximate}
Forsell, N., and Sabbadin, R., \enquote{Approximate linear-programming
  algorithms for graph-based Markov decision processes,} \emph{Proceedings of
  the 2006 conference on ECAI 2006: 17th European Conference on Artificial
  Intelligence August 29--September 1, 2006, Riva del Garda, Italy}, 2006, pp.
  590--594.

\bibitem[{Cheng et~al.(2013)Cheng, Liu, Chen, and Ihler}]{cheng2013variational}
Cheng, Q., Liu, Q., Chen, F., and Ihler, A.~T., \enquote{Variational planning
  for graph-based MDPs,} \emph{Advances in Neural Information Processing
  Systems}, Vol.~26, 2013, pp. 2976--2984.

\bibitem[{Blondel and Tsitsiklis(2000)}]{blondel2000survey}
Blondel, V.~D., and Tsitsiklis, J.~N., \enquote{A survey of computational
  complexity results in systems and control,} \emph{Automatica}, Vol.~36,
  No.~9, 2000, pp. 1249--1274.

\bibitem[{Hernandez-Leal et~al.(2019)Hernandez-Leal, Kartal, and
  Taylor}]{hernandez2019survey}
Hernandez-Leal, P., Kartal, B., and Taylor, M.~E., \enquote{A survey and
  critique of multiagent deep reinforcement learning,} \emph{Autonomous Agents
  and Multi-Agent Systems}, Vol.~33, No.~6, 2019, pp. 750--797.

\bibitem[{Littman(1994)}]{littman1994markov}
Littman, M.~L., \enquote{Markov games as a framework for multi-agent
  reinforcement learning,} \emph{Machine learning proceedings 1994}, Elsevier,
  1994, pp. 157--163.

\bibitem[{Claus and Boutilier(1998)}]{claus1998dynamics}
Claus, C., and Boutilier, C., \enquote{The dynamics of reinforcement learning
  in cooperative multiagent systems,} \emph{AAAI/IAAI}, Vol. 1998, No. 746-752,
  1998, p.~2.

\bibitem[{Littman(2001)}]{littman2001value}
Littman, M.~L., \enquote{Value-function reinforcement learning in Markov
  games,} \emph{Cognitive systems research}, Vol.~2, No.~1, 2001, pp. 55--66.

\bibitem[{Hu and Wellman(2003)}]{hu2003nash}
Hu, J., and Wellman, M.~P., \enquote{Nash Q-learning for general-sum stochastic
  games,} \emph{Journal of machine learning research}, Vol.~4, No. Nov, 2003,
  pp. 1039--1069.

\bibitem[{Silver et~al.(1990)Silver, Frawley, Iba, Vittal, and
  Bradford}]{silver1990ils}
Silver, B., Frawley, W., Iba, G., Vittal, J., and Bradford, K., \enquote{ILS: A
  framework for multi-paradigmatic learning,} \emph{Machine Learning
  Proceedings 1990}, Elsevier, 1990, pp. 348--356.

\bibitem[{Hu and Liu(2020)}]{hu2020uas}
Hu, J., and Liu, Y., \enquote{UAS Conflict Resolution Integrating a Risk-Based
  Operational Safety Bound as Airspace Reservation with Reinforcement
  Learning,} \emph{AIAA Scitech 2020 Forum}, 2020, p. 1372.

\bibitem[{Wang et~al.(2021)Wang, Liu, Srikant, and Ying}]{wang20213m}
Wang, W., Liu, Y., Srikant, R., and Ying, L., \enquote{3M-RL: Multi-Resolution,
  Multi-Agent, Mean-Field Reinforcement Learning for Autonomous UAV Routing,}
  \emph{IEEE Transactions on Intelligent Transportation Systems}, 2021.

\bibitem[{Zhang et~al.(2018)Zhang, Yang, Liu, Zhang, and
  Basar}]{zhang2018fully}
Zhang, K., Yang, Z., Liu, H., Zhang, T., and Basar, T., \enquote{Fully
  decentralized multi-agent reinforcement learning with networked agents,}
  \emph{International Conference on Machine Learning}, PMLR, 2018, pp.
  5872--5881.

\bibitem[{Guestrin et~al.(2001)Guestrin, Koller, and
  Parr}]{guestrin2001multiagent}
Guestrin, C., Koller, D., and Parr, R., \enquote{Multiagent Planning with
  Factored MDPs.} \emph{NIPS}, Vol.~1, 2001, pp. 1523--1530.

\bibitem[{Cubuktepe et~al.(2021)Cubuktepe, Xu, and
  Topcu}]{cubuktepe2021distributed}
Cubuktepe, M., Xu, Z., and Topcu, U., \enquote{Distributed Policy Synthesis of
  Multi-Agent Systems with Graph Temporal Logic Specifications,} \emph{IEEE
  Transactions on Control of Network Systems}, 2021.

\bibitem[{Foerster et~al.(2018)Foerster, Farquhar, Afouras, Nardelli, and
  Whiteson}]{foerster2018counterfactual}
Foerster, J., Farquhar, G., Afouras, T., Nardelli, N., and Whiteson, S.,
  \enquote{Counterfactual multi-agent policy gradients,} \emph{Proceedings of
  the AAAI Conference on Artificial Intelligence}, Vol.~32, 2018.

\bibitem[{Lowe et~al.(2017)Lowe, Wu, Tamar, Harb, Abbeel, and
  Mordatch}]{lowe2017multi}
Lowe, R., Wu, Y., Tamar, A., Harb, J., Abbeel, P., and Mordatch, I.,
  \enquote{Multi-agent actor-critic for mixed cooperative-competitive
  environments,} \emph{arXiv preprint arXiv:1706.02275}, 2017.

\bibitem[{Qu et~al.(2020)Qu, Wierman, and Li}]{qu2020scalable}
Qu, G., Wierman, A., and Li, N., \enquote{Scalable reinforcement learning of
  localized policies for multi-agent networked systems,} \emph{Learning for
  Dynamics and Control}, PMLR, 2020, pp. 256--266.

\bibitem[{Icarte et~al.(2018)Icarte, Klassen, Valenzano, and
  McIlraith}]{icarte2018using}
Icarte, R.~T., Klassen, T., Valenzano, R., and McIlraith, S., \enquote{Using
  reward machines for high-level task specification and decomposition in
  reinforcement learning,} \emph{International Conference on Machine Learning},
  PMLR, 2018, pp. 2107--2116.

\bibitem[{Neary et~al.(2020)Neary, Xu, Wu, and Topcu}]{neary2020reward}
Neary, C., Xu, Z., Wu, B., and Topcu, U., \enquote{Reward machines for
  cooperative multi-agent reinforcement learning,} \emph{arXiv preprint
  arXiv:2007.01962}, 2020.

\bibitem[{DeFazio and Zhang(2021)}]{defazio2021learning}
DeFazio, D., and Zhang, S., \enquote{Learning Quadruped Locomotion Policies
  with Reward Machines,} \emph{arXiv preprint arXiv:2107.10969}, 2021.

\bibitem[{Shah et~al.(2020)Shah, Li, and Shah}]{shah2020planning}
Shah, A., Li, S., and Shah, J., \enquote{Planning with uncertain specifications
  (puns),} \emph{IEEE Robotics and Automation Letters}, Vol.~5, No.~2, 2020,
  pp. 3414--3421.

\bibitem[{Xu et~al.(2020)Xu, Gavran, Ahmad, Majumdar, Neider, Topcu, and
  Wu}]{xu2020joint}
Xu, Z., Gavran, I., Ahmad, Y., Majumdar, R., Neider, D., Topcu, U., and Wu, B.,
  \enquote{Joint inference of reward machines and policies for reinforcement
  learning,} \emph{Proceedings of the International Conference on Automated
  Planning and Scheduling}, Vol.~30, 2020, pp. 590--598.

\bibitem[{Toro~Icarte et~al.(2019)Toro~Icarte, Waldie, Klassen, Valenzano,
  Castro, and McIlraith}]{toro2019learning}
Toro~Icarte, R., Waldie, E., Klassen, T., Valenzano, R., Castro, M., and
  McIlraith, S., \enquote{Learning reward machines for partially observable
  reinforcement learning,} \emph{Advances in Neural Information Processing
  Systems}, Vol.~32, 2019, pp. 15523--15534.

\bibitem[{Furelos-Blanco et~al.(2020)Furelos-Blanco, Law, Russo, Broda, and
  Jonsson}]{furelos2020induction}
Furelos-Blanco, D., Law, M., Russo, A., Broda, K., and Jonsson, A.,
  \enquote{Induction of subgoal automata for reinforcement learning,}
  \emph{Proceedings of the AAAI Conference on Artificial Intelligence},
  Vol.~34, 2020, pp. 3890--3897.

\bibitem[{Furelos-Blanco et~al.(2021)Furelos-Blanco, Law, Jonsson, Broda, and
  Russo}]{furelos2021induction}
Furelos-Blanco, D., Law, M., Jonsson, A., Broda, K., and Russo, A.,
  \enquote{Induction and exploitation of subgoal automata for reinforcement
  learning,} \emph{Journal of Artificial Intelligence Research}, Vol.~70, 2021,
  pp. 1031--1116.

\bibitem[{Hasanbeig et~al.(2021)Hasanbeig, Jeppu, Abate, Melham, and
  Kroening}]{hasanbeig2021deepsynth}
Hasanbeig, M., Jeppu, N.~Y., Abate, A., Melham, T., and Kroening, D.,
  \enquote{DeepSynth: Automata Synthesis for Automatic Task Segmentation in
  Deep Reinforcement Learning,} \emph{The Thirty-Fifth $\{$AAAI$\}$ Conference
  on Artificial Intelligence,$\{$AAAI$\}$}, Vol.~2, Baltimore, MD, USA, August
  15--17, 27th $\{$USENIX$\}$ Security Symposium,$\{$USENIX$\}$ Security 18,
  2021, p.~36.

\bibitem[{Rens and Raskin(2020)}]{rens2020learning}
Rens, G., and Raskin, J.-F., \enquote{Learning non-markovian reward models in
  mdps,} \emph{arXiv preprint arXiv:2001.09293}, 2020.

\bibitem[{Parr and Russell(1998)}]{parr1998reinforcement}
Parr, R., and Russell, S., \enquote{Reinforcement learning with hierarchies of
  machines,} \emph{Advances in neural information processing systems}, 1998,
  pp. 1043--1049.

\bibitem[{Sutton et~al.(1999)Sutton, Precup, and Singh}]{sutton1999between}
Sutton, R.~S., Precup, D., and Singh, S., \enquote{Between MDPs and semi-MDPs:
  A framework for temporal abstraction in reinforcement learning,}
  \emph{Artificial intelligence}, Vol. 112, No. 1-2, 1999, pp. 181--211.

\bibitem[{Dietterich(2000)}]{dietterich2000hierarchical}
Dietterich, T.~G., \enquote{Hierarchical reinforcement learning with the MAXQ
  value function decomposition,} \emph{Journal of artificial intelligence
  research}, Vol.~13, 2000, pp. 227--303.

\bibitem[{Icarte et~al.(2020)Icarte, Klassen, Valenzano, and
  McIlraith}]{icarte2020reward}
Icarte, R.~T., Klassen, T.~Q., Valenzano, R., and McIlraith, S.~A.,
  \enquote{Reward Machines: Exploiting Reward Function Structure in
  Reinforcement Learning,} \emph{arXiv preprint arXiv:2010.03950}, 2020.

\bibitem[{Sutton et~al.(2000)Sutton, McAllester, Singh, and
  Mansour}]{sutton2000policy}
Sutton, R.~S., McAllester, D.~A., Singh, S.~P., and Mansour, Y.,
  \enquote{Policy gradient methods for reinforcement learning with function
  approximation,} \emph{Advances in neural information processing systems},
  2000, pp. 1057--1063.

\bibitem[{Della~Rossa et~al.(2020)Della~Rossa, Salzano, Di~Meglio, De~Lellis,
  Coraggio, Calabrese, Guarino, Cardona, DeLellis, Liuzza
  et~al.}]{della2020intermittent}
Della~Rossa, F., Salzano, D., Di~Meglio, A., De~Lellis, F., Coraggio, M.,
  Calabrese, C., Guarino, A., Cardona, R., DeLellis, P., Liuzza, D., et~al.,
  \enquote{Intermittent yet coordinated regional strategies can alleviate the
  COVID-19 epidemic: a network model of the Italian case,} \emph{arXiv preprint
  arXiv:2005.07594}, 2020.

\bibitem[{Sutton and Barto(2018)}]{sutton2018reinforcement}
Sutton, R.~S., and Barto, A.~G., \emph{Reinforcement learning: An
  introduction}, MIT press, 2018.

\bibitem[{Miles et~al.(2021)Miles, Stedman, and Heald}]{miles2021stay}
Miles, D.~K., Stedman, M., and Heald, A.~H., \enquote{“Stay at Home, Protect
  the National Health Service, Save Lives”: a cost benefit analysis of the
  lockdown in the United Kingdom,} \emph{International Journal of Clinical
  Practice}, Vol.~75, No.~3, 2021, p. e13674.

\bibitem[{Tan(1993)}]{tan1993multi}
Tan, M., \enquote{Multi-agent reinforcement learning: Independent vs.
  cooperative agents,} \emph{Proceedings of the tenth international conference
  on machine learning}, 1993, pp. 330--337.

\bibitem[{Schulman et~al.(2017)Schulman, Wolski, Dhariwal, Radford, and
  Klimov}]{schulman2017proximal}
Schulman, J., Wolski, F., Dhariwal, P., Radford, A., and Klimov, O.,
  \enquote{Proximal policy optimization algorithms,} \emph{arXiv preprint
  arXiv:1707.06347}, 2017.

\bibitem[{Andrychowicz et~al.(2017)Andrychowicz, Wolski, Ray, Schneider, Fong,
  Welinder, McGrew, Tobin, Abbeel, and Zaremba}]{andrychowicz2017hindsight}
Andrychowicz, M., Wolski, F., Ray, A., Schneider, J., Fong, R., Welinder, P.,
  McGrew, B., Tobin, J., Abbeel, P., and Zaremba, W., \enquote{Hindsight
  experience replay,} \emph{arXiv preprint arXiv:1707.01495}, 2017.

\end{thebibliography}

\newpage
\section{APPENDIX}

\subsection{Proof of theorem 1}

Recall $s = (s_{N_i^\kappa}, s_{N_{-i}^\kappa}), u = (u_{N_i^\kappa}, u_{N_{-i}^\kappa}), a = (a_{N_i^\kappa}, a_{N_{-i}^\kappa})$. We denote $s' = (s_{N_i^\kappa}, s_{N_{-i}^\kappa}'),u' = (u_{N_i^\kappa}, u_{N_{-i}^\kappa}'), a' = (a_{N_i^\kappa}, a_{N_{-i}^\kappa}')$. $d_{t, i}$ is the distribution of $(s_i(t), u_i(t), a_i(t))$ conditioned on $(s(0), u(0), a(0)) = (s, u, a)$ under policy $\theta$, and $d_{t, i}'$ is the distribution of $(s_i(t), u_i(t), a_i(t))$ conditioned on $(s(0), u(0), a(0)) = (s', u', a')$ under policy $\theta$. Since the system transition model $P_{G \mathcal{R}}$ is local dependent and the policy $\pi_i^{\theta_i}$ is localized, $d_{t, i}$ only depends on the initial information of agent $i$'s $t$-hop neighborhood, $(s_{N_i^t}, u_{N_i^t}, a_{N_i^t})$. Thus, $d_{t, i}'$ = $d_{t, i}$, for all $t \le \kappa$. Expanding  $Q_i^\theta$ using Eq. 4, we can get
\begin{equation}
    \begin{aligned}
        & |Q_i^\theta(s,u,a) - Q_i^\theta(s',u',a')| \\
        & \le \sum_{t=0}^\infty \Big |\mathbb{E}\big[\gamma^t R_i(s_i(t), u_i(t), a_i(t)|(s(0), u(0), a(0)) = (s,u, a)\big] \\
        & \quad - \mathbb{E}\big[\gamma^t R_i(s_i(t), u_i(t), a_i(t)|(s(0), u(0), a(0)) = (s', u', a')\big] \Big | \\
        & \le \sum_{t=0}^\infty \Big|\gamma^t \mathbb{E}_{(s_i, u_i, a_i) \sim d_{t,i}}R_i(s_i, u_i, a_i) - \gamma^t \mathbb{E}_{(s_i, u_i, a_i) \sim d_{t,i}'}R_i(s_i, u_i, a_i)\Big|\\
        &= \sum_{t=\kappa+1}^\infty \Big |\gamma^t \mathbb{E}_{(s_i, u_i, a_i) \sim d_{t,i}}R_i(s_i, u_i, a_i)  - \gamma^t \mathbb{E}_{(s_i, u_i, a_i) \sim d_{t,i}'}R_i(s_i, u_i, a_i)\Big| \\
        & \le \sum_{t=\kappa+1}^\infty \gamma^t \overline{R} TV(d_{t,i}, d_{t,i}') \\
        & \le \frac{\overline{R}}{1-\gamma} \gamma^{\kappa+1}
    \end{aligned} 
\end{equation}
where $TV(d_{t,i}, d_{t,i}')$ is the total variation distance between $d_{t,i}$ and $d_{t,i}'$, which is no larger than 1.

\subsection{Proof of theorem 2}
For any $(s, u, a) \in S_G \times U_G \times A_G$,
\begin{equation}
\begin{aligned}
    & |\Tilde{Q}_i^\theta(s_{N_i^\kappa}, u_{N_i^\kappa}, a_{N_i^\kappa})-Q_i^\theta(s, u,a)| \\
    & \le \big|\sum_{s_{N_{-i}^\kappa}', u_{N_{-i}^\kappa}', a_{N_{-i}^\kappa}'}
    c_i(s_{N_{-i}^\kappa}',u_{N_{-i}^\kappa}',a_{N_{-i}^\kappa}'; s_{N_i^\kappa},u_{N_i^\kappa},a_{N_i^\kappa})  Q_i^\theta(s_{N_i^\kappa},s_{N_{-i}^\kappa}',u_{N_i^\kappa},u_{N_{-i}^\kappa}', a_{N_i^\kappa},a_{N_{-i}^\kappa}') \\
    &\quad - Q_i^\theta(s_{N_i^\kappa},s_{N_{-i}^\kappa},u_{N_i^\kappa},u_{N_{-i}^\kappa}, a_{N_i^\kappa},a_{N_{-i}^\kappa}) \big| \\
    & \le  \sum_{s_{N_{-i}^\kappa}', u_{N_{-i}^\kappa}', a_{N_{-i}^\kappa}'}
    c_i(s_{N_{-i}^\kappa}',u_{N_{-i}^\kappa}',a_{N_{-i}^\kappa}'; s_{N_i^\kappa},u_{N_i^\kappa},a_{N_i^\kappa}) \\
    &  \quad \cdot \big |Q_i^\theta(s_{N_i^\kappa},s_{N_{-i}^\kappa}',u_{N_i^\kappa},u_{N_{-i}^\kappa}', a_{N_i^\kappa},a_{N_{-i}^\kappa}')  - Q_i^\theta(s_{N_i^\kappa},s_{N_{-i}^\kappa},u_{N_i^\kappa},u_{N_{-i}^\kappa}, a_{N_i^\kappa},a_{N_{-i}^\kappa}) \big | \\
    & \le \lambda \rho^{\kappa+1}
\end{aligned}
\end{equation}

\subsection{Proof of theorem 3}
Since $\pi^\theta(a|s, u) = \prod_{i=1}^N \pi_i^{\theta_i} (a_i|s_i, u_i)$, we have $\nabla_{\theta_i} \log \pi^\theta(a|s,u) = \nabla_{\theta_i} \sum_{j \in V} \log \pi_j^{\theta_j}(a_j|s_j, u_j) = \nabla_{\theta_i} \log \pi_i^{\theta_i}(a_i|s_i, u_i)$. Therefore,
\begin{equation}
\begin{aligned}
    \nabla_{\theta_i} J(\theta) &= \frac{1}{1-\gamma} \mathbb{E}_{(s,u) \sim d^\theta, a \sim \pi^\theta(\cdot |s,u)} Q^\theta(s,u,a)  \nabla_{\theta_i} \log \pi^\theta(a|s,u) \\
    &= \frac{1}{1-\gamma} \mathbb{E}_{(s,u) \sim d^\theta, a \sim \pi^\theta(\cdot |s,u)} Q^\theta(s,u,a) \nabla_{\theta_i} \log \pi_i^{\theta_i}(a_i|s_i,u_i)
\end{aligned}
\end{equation}
Recall $\Tilde{J}_i(\theta) = \frac{1}{1-\gamma} \mathbb{E}_{(s,u) \sim d^\theta, a \sim \pi^\theta(\cdot |s,u)} [\frac{1}{n} \sum_{j \in N_i^\kappa} \Tilde{Q}_j^\theta(s_{N_j^\kappa}, u_{N_j^\kappa}, a_{N_j^\kappa})] \nabla_{\theta_i} \log \pi_i^{\theta_i}(a_i|s_i. u_i)$. Then, we compute $\Tilde{J}_i(\theta) - \nabla_{\theta_i} J(\theta)$,
\begin{equation}
\begin{aligned}
    & \Tilde{J}_i(\theta) - \nabla_{\theta_i} J(\theta) \\
    &= \frac{1}{1-\gamma} \mathbb{E}_{(s,u) \sim d^\theta, a \sim \pi^\theta(\cdot |s,u)} \big [\frac{1}{n} \sum_{j \in N_i^\kappa} \Tilde{Q}_j^\theta(s_{N_j^\kappa}, u_{N_j^\kappa}, a_{N_j^\kappa}) - Q^\theta(s,u,a) \big]  \nabla_{\theta_i} \log \pi_i^{\theta_i}(a_i|s_i,u_i) \\
    &= \frac{1}{1-\gamma} \mathbb{E}_{(s,u) \sim d^\theta, a \sim \pi^\theta(\cdot |s,u)} \big[\frac{1}{n} \sum_{j \in N_i^\kappa} \Tilde{Q}_j^\theta(s_{N_j^\kappa}, u_{N_j^\kappa}, a_{N_j^\kappa}) - \frac{1}{n} \sum_{j \in V}Q_j^\theta(s,u,a)\big]  \nabla_{\theta_i} \log \pi_i^{\theta_i}(a_i|s_i,u_i) \\
    & \quad -  \frac{1}{1-\gamma} \mathbb{E}_{(s,u) \sim d^\theta, a \sim \pi^\theta(\cdot |s,u)} \frac{1}{n} \sum_{j \in N_{-i}^\kappa} \Tilde{Q}_j^\theta(s_{N_j^\kappa}, u_{N_j^\kappa}, a_{N_j^\kappa})\nabla_{\theta_i} \log \pi_i^{\theta_i}(a_i|s_i,u_i)\\
    &= e_1 - e_2
\end{aligned}
\end{equation}
We first show $e_2 = 0$. Let $e_3 = \mathbb{E}_{(s,u) \sim d^\theta, a \sim \pi^\theta(\cdot |s,u)} \nabla_{\theta_i} \log \pi_i^{\theta_i} (a_i|s_i,u_i) \Tilde{Q}_j^\theta(s_{N_j^\kappa}, u_{N_j^\kappa}, a_{N_j^\kappa})$, for any $j \in N_j^\kappa$,
\begin{equation}
\begin{aligned}
    e_3 &= \sum_{s, u, a} d^\theta(s, u) \prod_{\overline{i}=1}^n \pi_{\overline{i}}^{\theta_{\overline{i}}}(a_{\overline{i}} | s_{\overline{i}}, u_{\overline{i}}) \frac{\nabla_{\theta_i} \pi_i^{\theta_i} (a_i|s_i, u_i)}{\pi_i^{\theta_i} (a_i|s_i, u_i)}  \Tilde{Q}_j^\theta(s_{N_j^\kappa}, u_{N_j^\kappa}, a_{N_j^\kappa}) \\
    &= \sum_{s, u, a} d^\theta(s, u) \prod_{\overline{i} \ne i} \pi_{\overline{i}}^{\theta_{\overline{i}}}(a_{\overline{i}} | s_{\overline{i}}, u_{\overline{i}}) \nabla_{\theta_i} \pi_i^{\theta_i} (a_i|s_i, u_i) \Tilde{Q}_j^\theta(s_{N_j^\kappa}, u_{N_j^\kappa}, a_{N_j^\kappa}) \\
    &= \sum_{s,u,a_1, \dots, a_{i-1}, a_{i+1}, \dots, a_n} d^\theta(s,u ) \prod_{\overline{i} \ne i} \pi_{\overline{i}}^{\theta_{\overline{i}}}(a_{\overline{i}} | s_{\overline{i}}, u_{\overline{i}}) \Tilde{Q}_j^\theta(s_{N_j^\kappa}, u_{N_j^\kappa}, a_{N_j^\kappa}) \sum_{a_i} \nabla_{\theta_i} \pi_i^{\theta_i} (a_i|s_i, u_i) \\
    &= \sum_{s,u,a_1, \dots, a_{i-1}, a_{i+1}, \dots, a_n} d^\theta(s,u ) \prod_{\overline{i} \ne i} \pi_{\overline{i}}^{\theta_{\overline{i}}}(a_{\overline{i}} | s_{\overline{i}}, u_{\overline{i}})   \Tilde{Q}_j^\theta(s_{N_j^\kappa}, u_{N_j^\kappa}, a_{N_j^\kappa}) \nabla_{\theta_i} \sum_{a_i}  \pi_i^{\theta_i} (a_i|s_i, u_i)\\
\end{aligned}
\end{equation}
Because $\Tilde{Q}_j^\theta(s_{N_j^\kappa}, u_{N_j^\kappa}, a_{N_j^\kappa})$ does not rely on $a_i$, as $i \notin N_j^\kappa$, and $\sum_{a_i} \nabla_{\theta_i} \pi_i^{\theta_i} (a_i|s_i, u_i) = \nabla_{\theta_i} \sum_{a_i}  \pi_i^{\theta_i} (a_i|s_i, u_i) =\nabla_{\theta_i} 1 = 0$, we can get $e_3 = 0, e_2 = 0$. Then we can have,
\begin{equation}
    \begin{aligned}
        & || \Tilde{J}_i(\theta) - \nabla_{\theta_i} J(\theta)||  \\
        & = ||e_1|| \\
        & \le \frac{1}{1-\gamma} \mathbb{E}_{(s,u) \sim d^\theta, a \sim \pi^\theta(\cdot |s,u)} \frac{1}{n} \sum_{j \in V} \big |\Tilde{Q}_j^\theta(s_{N_j^k},u_{N_j^k}, a_{N_j^k}) - Q_j^\theta(s, a) \big |  ||\nabla_{\theta_i} \log \pi_i^{\theta_i}(a_i|s_i, u_i)|| \\
        & \le \frac{1}{1-\gamma} \lambda \rho^{\kappa+1} B_i
    \end{aligned}
\end{equation}

\subsection{Details of Case study I: UAV Package Delivery}

\begin{figure}[H]
\centering
\includegraphics[width=0.45\columnwidth]{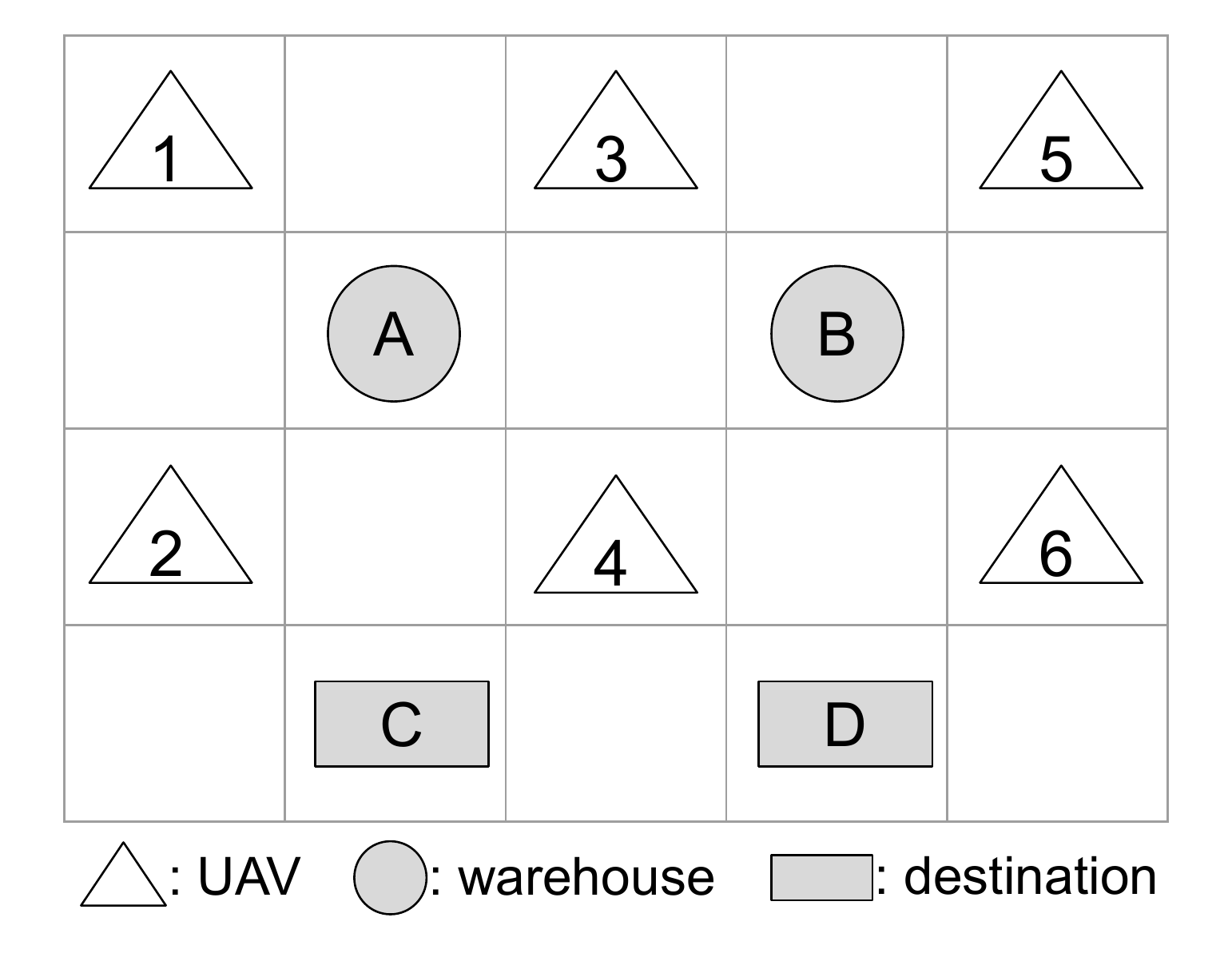}
\caption{Grid environment in case study I. Triangle with number $i$: initial position of UAV $i$.}.
\label{grid_env}
\end{figure}

The 4$\times$5 grid-world for this experiment is shown in \cref{grid_env}. Packages in warehouse A need to be transported to destination C, and packages in warehouse B need to be transported to destination D. UAV 1 and 2 only have access to warehouse A, UAV 5 and 6 only have access to warehouse B, and UAV 3 and 4 can go to either warehouse A or B. UAVs start from the positions as indicated in \cref{grid_env} with a fully charged battery. The battery consumption is $0.01\%$ when waiting and $2\%$ otherwise. We define a UAV runs out of battery if the remaining battery percentage is less than 7.5\%.

Regarding the reward machine for UAV $i\in \{3, 4\}$ as shown in Fig. 1b of the full paper, the labeled events for UAV $i\in \{3, 4\}$ are $\mathcal{P}_i = \{\hat{A}_{i1}, \hat{A}_{i2}, \hat{P}_1, \hat{P}_2, \hat{G}_{i1}, \hat{G}_{i2}, \hat{L} \}$, where the subscripts including 1 and 2 are related to the two accessible warehouses and the corresponding destinations. For instance, we let $A_{i1}$ mean UAV $i\in \{3, 4\}$ reaches warehouse A and $A_{i2}$ means arriving at warehouse B. Both $u_3$ and $u_6$ are goal states, and transitioning to either one indicates task completion.

\begin{figure}[H]
\centering
\includegraphics[width=0.9\textwidth]{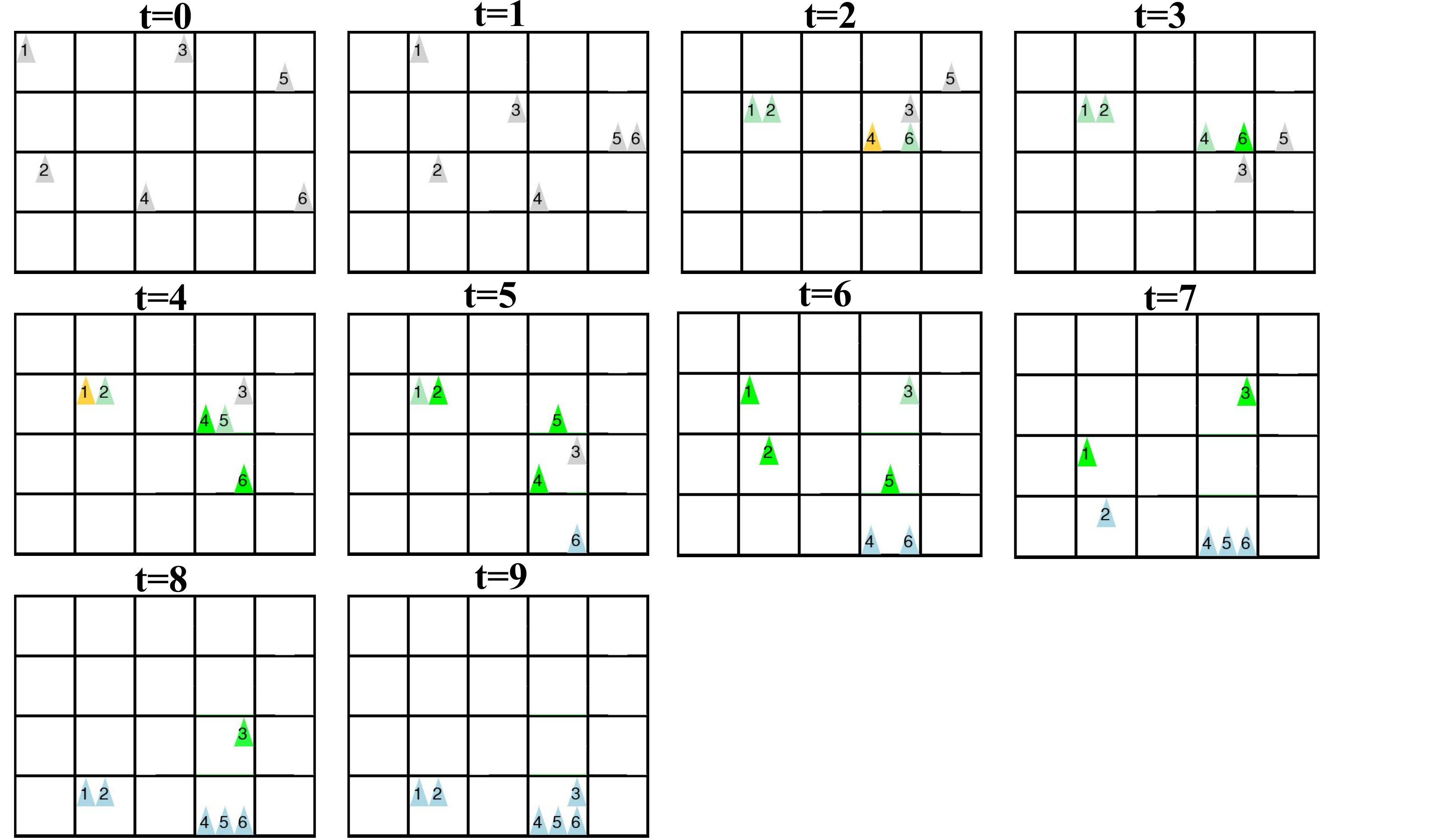}
\caption{Trajectories of UAVs with $\kappa=0$ in case study I. Yellow: select to wait; Light green: select to pick up; Dark green: receive a package; Blue: finish the task.}.
\label{grid_result}
\end{figure}

The trajectories of UAVs with $\kappa=0$ are shown in \cref{grid_result}. It can be seen  that all the UAVs finish the task.

\subsection{Details of Case study II: COVID-19 Pandemic Mitigation}

The bang-bang control of the baseline method is based on the relative saturation level of a region's health system, mathematically the ratio, $\Tilde{r_i}$, between the number of the hospitalized requiring care in ICU (estimated as $0.1\mathcal{H}_i$) over the number of available ICU beds ($\mathcal{T}_i^\mathcal{H}$) in the region. The regional lockdown is enforced when $\Tilde{r_i}$ is larger than 0.5 and relaxed when $\Tilde{r_i}$ is smaller than 0.2. During the lockdown, the region implements strict social distancing rules, and all fluxes in or out of the region are reduced to 70\% of their original values. Mathematically, the social distancing parameter and fluxes are set as follows.
\begin{equation}
    \begin{aligned}
        \rho_i &= \begin{cases}
                    \underline{\rho_i}   & \frac{0.1\mathcal{H}_i}{\mathcal{T}_i^\mathcal{H}} \ge 0.5  \\
                    \min(1, 3\underline{\rho_i})  & \frac{0.1\mathcal{H}_i}{\mathcal{T}_i^\mathcal{H}} \le 0.2
                  \end{cases} \\
    \end{aligned}
\end{equation}
where $\underline{\rho_i}$ is the minimum estimated value during the national lockdown and $\min(1, 3\underline{\rho_i})$ simulates the relaxation of social distancing rules.
\begin{equation}
    \begin{aligned}
       \phi_{ij} &= \begin{cases}
                    0.7\underline{\phi_{ij}}   & \frac{0.1\mathcal{H}_i}{\mathcal{T}_i^\mathcal{H}} \ge 0.5  \\
                    \underline{\phi_{ij}}  & \frac{0.1\mathcal{H}_i}{\mathcal{T}_i^\mathcal{H}} \le 0.2
                  \end{cases} , \forall i \ne j\\
        \phi_{ji} &= \begin{cases}
                    0.7\underline{\phi_{ji}}   & \frac{0.1\mathcal{H}_i}{\mathcal{T}_i^\mathcal{H}} \ge 0.5  \\
                    \underline{\phi_{ji}}  & \frac{0.1\mathcal{H}_i}{\mathcal{T}_i^\mathcal{H}} \le 0.2
                  \end{cases} , \forall i \ne j\\
        \phi_{ii} &= 1-\sum_{i\ne j}\phi_{ij}
    \end{aligned}
\end{equation}
where $\underline{\phi_{ij}}$ is the original value of the flux from region $i$ to region $j$ before the pandemic, and $0.7\underline{\phi_{ij}}$ simulates the feedback flux control during a lockdown. 

Recall $\Tilde{v}_i$ is the number of days when the region is in a severe situation since this Monday, $\Tilde{l}_i$ is the number of days when the region adopts a lockdown since this Monday, and the set of events is $\mathcal{P}_i=\{ \epsilon_0, \epsilon_1,  v^0l^0,  v^{0.5}l^0, v^1l^0, v^0l^{0.5},  v^{0.5}l^{0.5}, v^1l^{0.5}, \\ v^0l^1,  v^{0.5}l^1, v^1l^1\}$. Mathematically, 
\begin{equation}
    \begin{aligned}
        &v^k 
        = \begin{cases}
                v^0  &  \Tilde{v} = 0\\
                v^1  &  \Tilde{v} = 7\\
                v^{0.5} & 0 < \Tilde{v} < 7
                
           \end{cases} 
         &l^k 
        = \begin{cases}
                l^0  &  \Tilde{l} = 0\\
                l^1  &  \Tilde{l} = 7\\
                l^{0.5}  &  0 < \Tilde{l} < 7
                 \end{cases} \\
    \end{aligned}
    \label{eq:covid_reward}
\end{equation}

\begin{figure}[t]
\centering
\includegraphics[width=0.7\columnwidth]{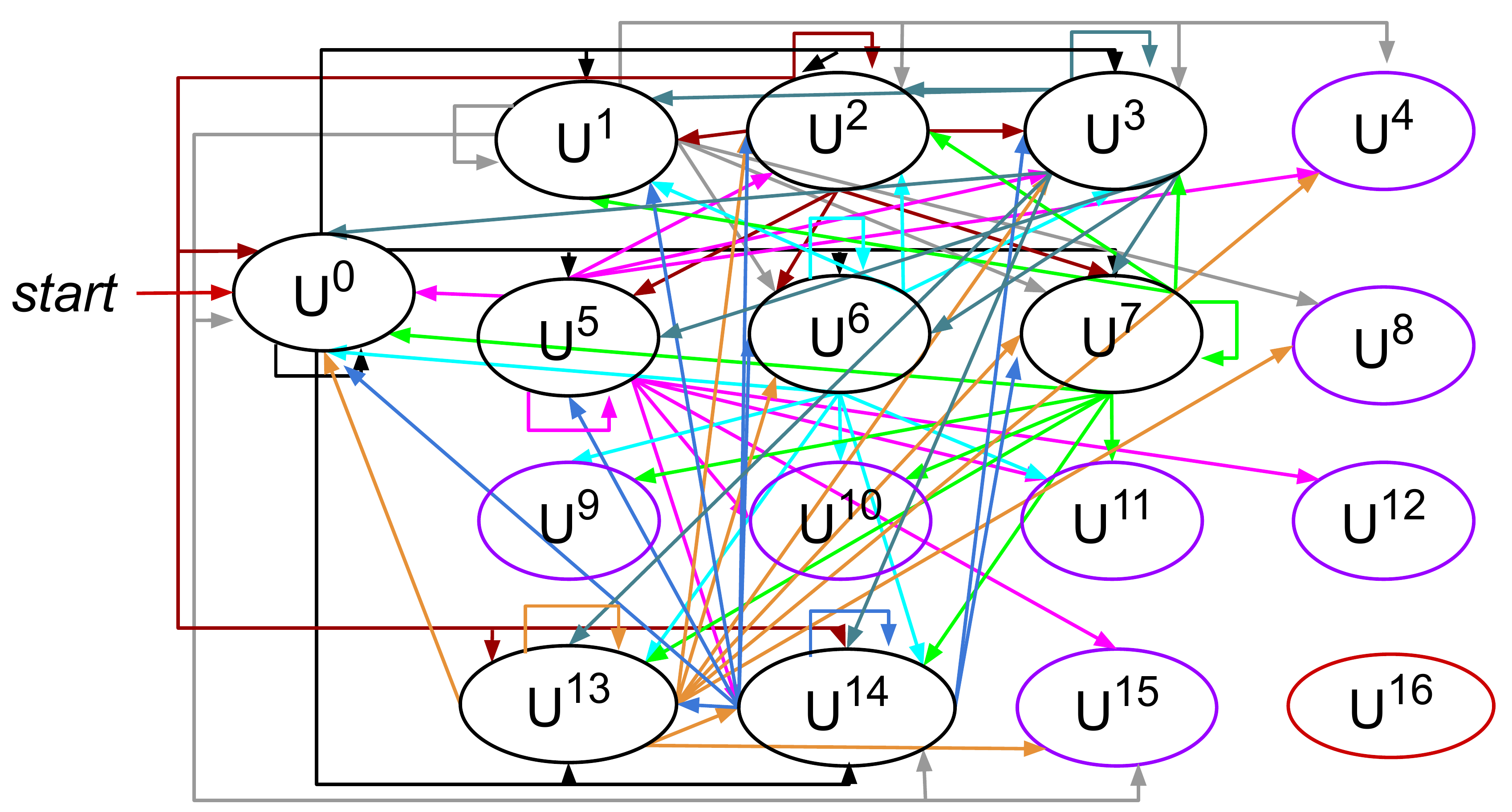}
\caption{Reward Machine for each agent in case study II. Purple: sink state; Red: goal state.}.
\label{RM_covid}
\end{figure}

The reward machine for task specification of each region is shown in \cref{RM_covid}. The agent starts at $u^0$, assuming during the previous week, the region was in a severe situation at least 1 day but less than 7 days and did not implement a lockdown. The RM states in purple ($u^4, u^8, u^9, u^{10}, u^{11}, u^{12}, u^{15}$) are sink states, indicating that the region has been in a severe situation or has implemented a lockdown for two consecutive weeks. $u^{16}$ is the goal state and can be reached from all the states except the sink states (arrows pointing to $u^{16}$ are omitted). Arrows with the same color point from the same RM state. Due to the limited space in \cref{RM_covid},  the required events for RM state transitions which should be marked on the edges in \cref{RM_covid} are listed in \cref{table:rm transition}. For instance, during the first week, if the region was in a severe situation and implemented a lockdown for seven consecutive days, denoted by $v^1l^1$, the region is directed to $u^5$ on the second week's Monday. From Tuesday to Sunday, the agent would remain at the same RM state due to the empty event $\epsilon_0$. The reward function $\sigma_i$ is defined as follows.
\begin{equation}
    \begin{aligned}
        \sigma_i(u_i, L_i(s_i, a_i, s'_i)) 
         = \begin{cases}
                -600  & u_i=u^4, u^8, u^9, u^{10}, u^{11}, u^{12}, u^{15} \\
                -250v-50l+400  & u'_i=u^3 \\
                -250v-50l-300  & u'_i=u^4, u^8, u^9, u^{10}, u^{11}, u^{12}, u^{15}\\
                -250v-50l+250  & L_i(s_i, a_i, s'_i) = \epsilon_1 \\
                0 & L_i(s_i, a_i, s'_i) = \epsilon_0 \\
                -250v-50l+200  & \text{otherwise} 
           \end{cases} \\
    \end{aligned}
    \label{eq:covid_reward}
\end{equation}
where $u'_i = \delta_i(u_i, L_i(s_i, a_i, s'_i))$.

\begin{table*}[]
    \centering
    \begin{tabular}{c|c|c|c|c|c|c|c|c|c}
    \hline
         $u$ transition & event &  $u$ transition & event & $u$ transition & event & $u$ transition & event & $u$ transition & event\\
    \hline
         $u_0 \to u_0$  & $\epsilon_0 \lor v^{0.5}l^0$ &  $u_0 \to u_1$ & $v^0l^1$ &  $u_0 \to u_2$ & $ v^0l^{0.5}$ & $u_0 \to u_3$ & $v^0l^0$ & $u_0 \to u_5$ & $ v^1l^1$\\
         $u_0 \to u_6$  & $ v^1l^{0.5}$ &  $u_0 \to u_7$ & $v^1l^0$ &  $u_0 \to u_{13}$ & $ v^{0.5}l^1$ & $u_0 \to u_{14}$ & $ v^{0.5}l^{0.5}$ & $u_0 \to u_{16}$ & $\epsilon_1$        \\
         
         $u_1 \to u_1$  & $\epsilon_0 \lor v^{0}l^1$ &  $u_1 \to u_2$ & $v^0l^{0.5}$ &  $u_1 \to u_3$ & $ v^0l^{0}$ & $u_1 \to u_4$ & $v^0l^1$ & $u_1 \to u_0$ & $ v^{0.5}l^0$\\
         $u_1 \to u_6$  & $ v^1l^{0.5}$ &  $u_1 \to u_7$ & $v^1l^0$ &  $u_1 \to u_8$ & $ v^{1}l^1$ & $u_1 \to u_{14}$ & $ v^{0.5}l^{0.5}$ & $u_1 \to u_{15}$ & $ v^{0.5}l^{1}$        \\
         $u_1 \to u_{16}$ & $\epsilon_1$ &  $u_2 \to u_2$ & $v^0l^{0.5}$ & $u_2 \to u_1$ & $v^{0}l^{1}$ & $u_2 \to u_0$ & $v^{0.5}l^{0}$ & $u_2 \to u_3$ & $ v^0l^{0}$ \\
         $u_2 \to u_5$ & $v^1l^1$ & $u_2 \to u_6$ & $ v^{1}l^{0.5}$ & $u_2 \to u_7$ & $v^1l^0$ & $u_2 \to u_13$ & $v^{0.5}l^1$ & $u_2 \to u_{14}$ & $v^{0.5}l^{0.5}$\\
         $u_2 \to u_{16}$ & $\epsilon_1$ & $u_3 \to u_3$ & $v^0l^{0}$ & $u_3 \to u_1$ & $v^{0}l^{1}$ & $u_3 \to u_2$ & $v^{0}l^{0.5}$ & $u_3 \to u_0$ & $ v^{0.5}l^{0}$\\ 
         $u_3 \to u_5$ & $v^1l^1$ & $u_3 \to u_6$ & $ v^{1}l^{0.5}$ & $u_3 \to u_7$ & $v^1l^0$ & $u_3 \to u_{13}$ & $v^{0.5}l^1$ & $u_3 \to u_{14}$ & $v^{0.5}l^{0.5}$\\
         $u_3 \to u_{16}$ & $\epsilon_1$ & $u_5 \to u_5$ & $v^1l^{1}$ & $u_5 \to u_0$ & $v^{0.5}l^{0}$ & $u_5 \to u_2$ & $v^{0}l^{0.5}$ & $u_5 \to u_3$ & $ v^{0}l^{0}$\\ 
         $u_5 \to u_4$ & $v^0l^1$ & $u_5 \to u_{10}$ & $ v^{1}l^{0.5}$ & $u_5 \to u_{11}$ & $v^1l^0$ & $u_5 \to u_{12}$ & $v^{1}l^1$ & $u_5 \to u_{14}$ & $v^{0.5}l^{0.5}$\\
         $u_5 \to u_{15}$ & $v^{0.5}l^{1}$ & $u_5 \to u_{16}$ & $\epsilon_1$ & $u_6 \to u_0$ & $v^{0.5}l^{0}$ & $u_6 \to u_2$ & $v^{0}l^{0.5}$ & $u_6 \to u_3$ & $ v^{0}l^{0}$ \\
         $u_6 \to u_6$ & $v^1l^{0.5}$ & $u_6 \to u_{10}$ & $ v^{1}l^{0.5}$ & $u_6 \to u_{11}$ & $v^1l^0$ & $u_6 \to u_{13}$ & $v^{0.5}l^1$ & $u_6 \to u_{14}$ & $v^{0.5}l^{0.5}$\\
         $u_6 \to u_1$ & $v^{0}l^{1}$ & $u_6 \to u_9$ & $v^{1}l^{1}$ & $u_6 \to u_{16}$ & $\epsilon_1$ & $u_7 \to u_7$ & $v^{1}l^{0}$ & $u_7 \to u_1$ & $v^{0}l^{1}$ \\
         $u_7 \to u_2$ & $v^{0}l^{0.5}$ & $u_7 \to u_3$ & $v^{0}l^{0}$ & $u_7 \to u_{0}$ & $v^{0.5}l^{0}$ & $u_7 \to u_9$ & $v^{1}l^{1}$ & $u_7 \to u_{10}$ & $v^{1}l^{0.5}$\\
         $u_7 \to u_{11}$ & $v^{1}l^{0}$ & $u_7 \to u_{13}$ & $v^{0.5}l^{1}$ & $u_7 \to u_{14}$ & $v^{0.5}l^{0.5}$ & $u_7 \to u_{16}$ & $\epsilon_1$ & $u_{13} \to u_{13}$ & $v^{0.5}l^{1}$\\
         $u_{13} \to u_{0}$ & $v^{0.5}l^{0}$ & $u_{13} \to u_{2}$ & $v^{0}l^{0.5}$ & $u_{13} \to u_{3}$ & $v^{0}l^{0}$ & $u_{13} \to u_{4}$ & $v^{0}l^{1}$ & $u_{13} \to u_{6}$ & $v^{1}l^{0.5}$\\
         $u_{13} \to u_{7}$ & $v^{1}l^{0}$ & $u_{13} \to u_{8}$ & $v^{1}l^{1}$ & $u_{13} \to u_{14}$ & $v^{0.5}l^{0.5}$ & $u_{13} \to u_{15}$ & $v^{0.5}l^{1}$ & $u_{13} \to u_{16}$ & $\epsilon_1$\\
         $u_{14} \to u_{14}$ & $v^{0.5}l^{0.5}$ & $u_{14} \to u_{13}$ & $v^{0.5}l^{1}$ & $u_{14} \to u_{0}$ & $v^{0.5}l^{0}$ & $u_{14} \to u_{1}$ & $v^{0}l^{1}$ & $u_{14} \to u_{2}$ & $v^{0}l^{0.5}$\\
         $u_{14} \to u_{3}$ & $v^{0}l^{0}$ & $u_{14} \to u_{5}$ & $v^{1}l^{1}$ & $u_{14} \to u_{6}$ & $v^{1}l^{0.5}$ & $u_{14} \to u_{7}$ & $v^{1}l^{0}$ & $u_{14} \to u_{16}$ & $\epsilon_1$\\

    \hline
    \end{tabular}
    \caption{Events for reward machine state transition in case study II.}
    \label{table:rm transition}
\end{table*}

\begin{figure}[H]
\centering
\includegraphics[width=0.6\columnwidth]{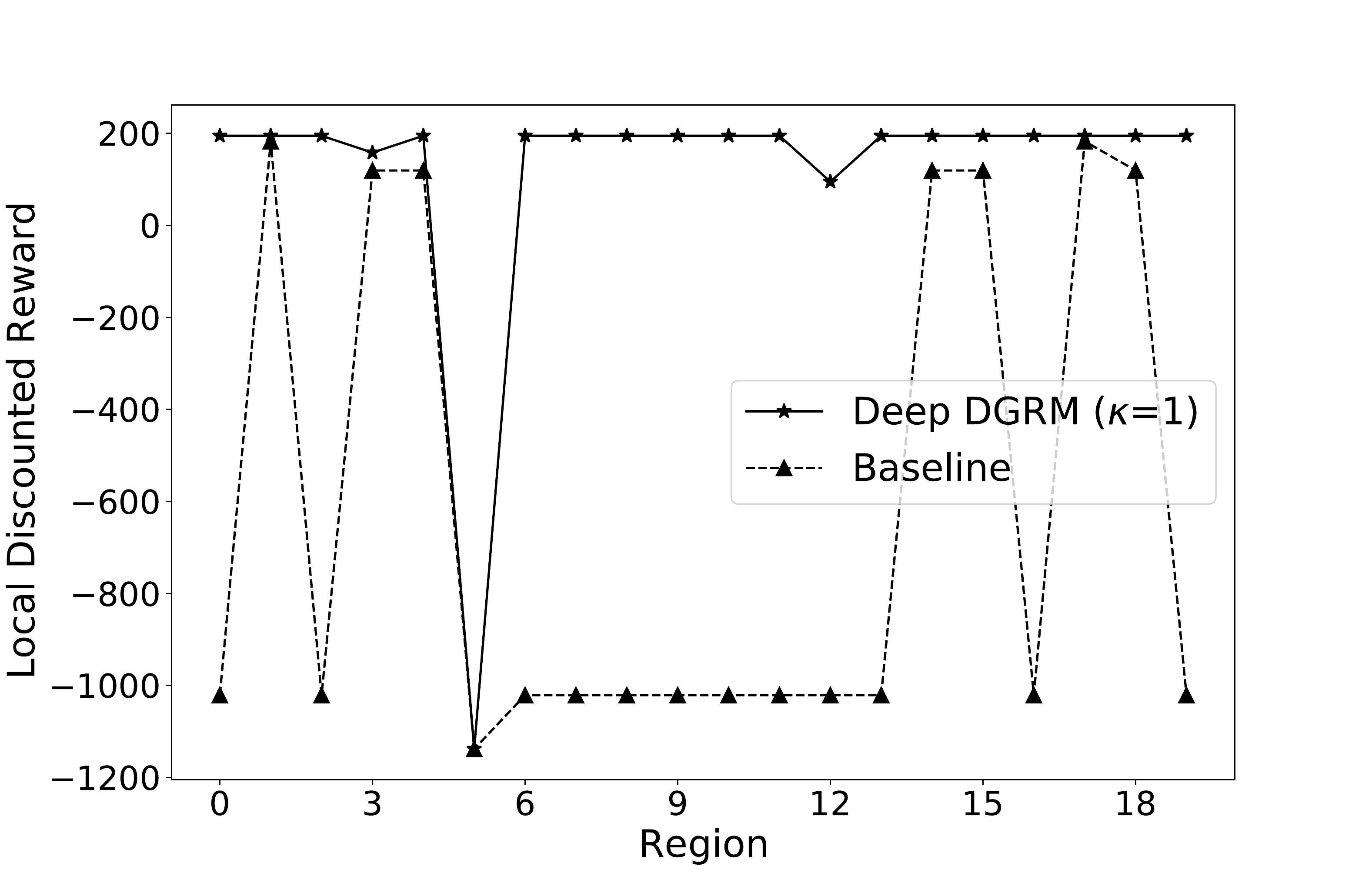}
\caption{Comparison of local discounted rewards of each region in case study II.}.
\label{covid_r_compare}
\end{figure}

\begin{figure}[H]
\centering
\includegraphics[width=0.96\textwidth]{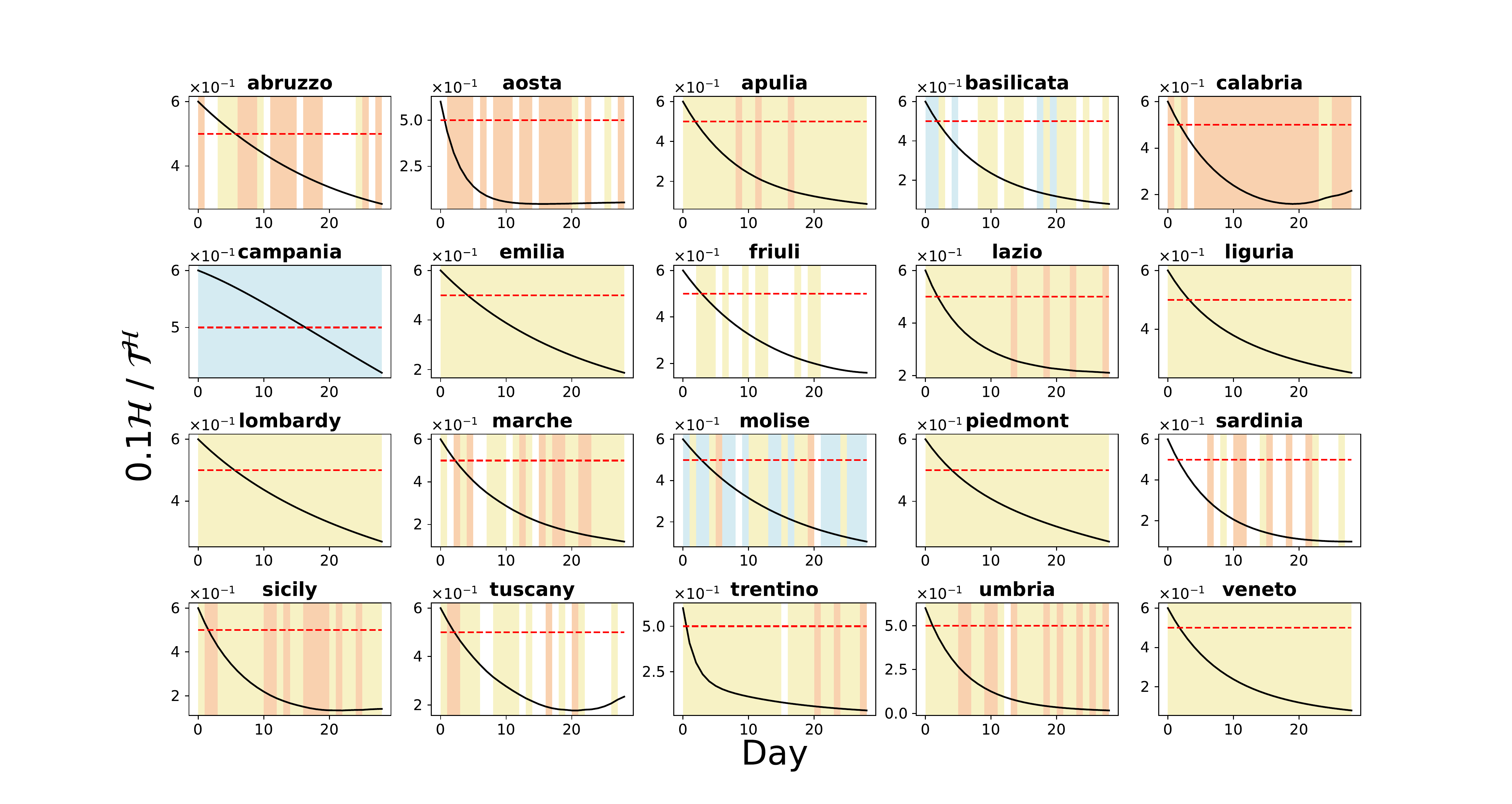}
\caption{Result of $\frac{0.1\mathcal{H}_i}{\mathcal{T}_i^\mathcal{H}}$ of each region with $\kappa = 1$ in case study II. Red dashed line: the threshold of 0.5 for  $\frac{0.1\mathcal{H}_i}{\mathcal{T}_i^\mathcal{H}}$ when the region is in a severe condition. Background: blue: select to lockdown; yellow: select to practice social distancing; orange: select to control the flux.}
\label{covid_city_h}
\end{figure}

\cref{covid_r_compare} shows the local discounted rewards of each region using the deep DRGM algorithm and the baseline method, with the reward function determined by the reward machine. From \cref{covid_r_compare}, it can be seen that only region 5 receives a large penalty and fails the task with the deep \foo ABC algorithm, while a total of 13 regions fail the task utilizing the baseline. We also plot the result of $\frac{0.1\mathcal{H}_i}{\mathcal{T}_i^\mathcal{H}}$ during the entire simulation for each region with $\kappa = 1$ in \cref{covid_city_h}. The result indicates our proposed deep \foo ABC algorithm can guide 19 among 20 regions to avoid being in a severe situation and adopting a lockdown for 2 consecutive weeks in a period of 4 weeks, which reduces the spread of disease while considering the economic factors.

\end{document}